\RequirePackage{fix-cm}
\documentclass[smallcondensed,natbib]{svjour3}     % onecolumn (ditto)
\smartqed  % flush right qed marks, e.g. at end of proof
\usepackage{amssymb}
\usepackage{amsmath}
\usepackage{amssymb}
\usepackage{graphicx}
\usepackage{longtable}
 \journalname{}
\begin{document}

\title{Analysis of DevR regulated genes in \textit{Mycobacterium tuberculosis}
%Mathematical modeling and information theoretical analysis of DevR regulated 
%genes in \textit{Mycobacterium tuberculosis}
}

%\titlerunning{Mathematical modeling $\ldots$ genes in 
%\textit{M. tuberculosis}
%}        % if too long for running head

\author{Arnab Bandyopadhyay 
              \and Soumi Biswas \\
              \and Alok Kumar Maity
              \and Suman K Banik$^*$
}

\authorrunning{Arnab Bandyopadhyay et al.} % if too long for running head

\institute{
Arnab Bandyopadhyay \and Soumi Biswas 
\and Suman K Banik (\email{skbanik@jcbose.ac.in})
\at Department of Chemistry, Bose Institute, 
93/1 A P C Road, Kolkata 700009, India. \\
Tel.: +91-33-2303-1142, Fax: +91-33-2303-6790 \\
Alok Kumar Maity
\at Department of Chemistry, University of Calcutta, 
92 A P C Road, Kolkata 700009, India. \\
$^*$Corresponding author
}

\date{Received: date / Accepted: date}

\maketitle

\begin{abstract}
The DevRS two component system of \textit{Mycobacterium tuberculosis} is responsible 
for its dormancy in host and becomes operative under hypoxic condition. It is experimentally
known that phosphorylated DevR controls the expression of several downstream genes in a 
complex manner. In the present work we propose a theoretical model to show 
role of binding sites in DevR mediated gene expression. Individual and collective role of
binding sites in regulating DevR mediated gene expression has been shown via modeling.
Objective of the present work is two fold. First, to describe qualitatively the temporal dynamics 
of wild type genes and their known mutants. Based on these results we propose that DevR 
controlled gene expression follows a specific pattern which is efficient in describing other 
DevR mediated gene expression. Second, to analyze behavior of the system from information 
theoretical point of view. Using the tools of information theory we have calculated molecular 
efficiency of the system and have shown that it is close to the maximum limit of isothermal 
efficiency.
\end{abstract}

%%%%%%%%%% Introduction

\section{Introduction}

\begin{figure}[!t]
\includegraphics[width=0.75\linewidth,angle=0]{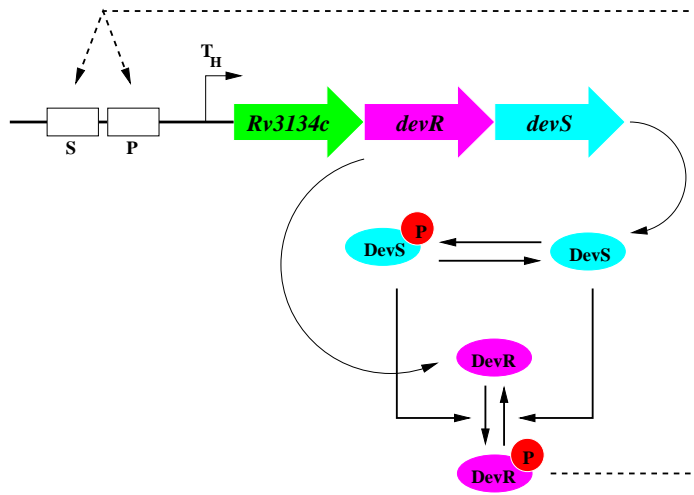}
\caption{Schematic diagram of signal transduction pathway in DevRS two component 
system. Positive feedback of phosphorylated DevR on its own operon and on
Rv3134c is shown by the dotted line. Two DevR binding sites S (distal) and P (proximal) 
are denoted by open boxes. $T_H$ denotes hypoxia inducible promoter for Rv3134c. 
For simplicity, we do not show mRNA and degradation of proteins in the diagram.
}
\label{network}
\end{figure}

\textit{Mycobacterium tuberculosis} is one of the most well studied human pathogen that causes 
around 2 million deaths each year. Persistency of  \textit{M. tuberculosis} in human body, 
sometimes for decades, makes it most deadly compared to other human pathogens. While 
residing within the human body \textit{M. tuberculosis} experiences different kind of stresses 
and/or signals in the form of chemical components. Most of these signals are sensed by the well 
defined two component systems (TCS). In order to respond to different environmental stimuli 
\textit{M. tuberculosis} has developed 11 well defined TCS \citep{Bretl2011} among which 
DevRS is one of  the most studied one and is particularly responsible for dormancy of 
\textit{M. tuberculosis} in host. Likewise other TCS in bacteria
\citep{Appleby1996,Bijlsma2003,Cotter2003,Hoch2000,Laub2007},  
DevRS is comprised of membrane bound sensor kinase 
DevS and cytoplasmic response regulator DevR. DevRS TCS becomes active under hypoxic, 
nitric oxide or nutrient starvation conditions through autophosphorylation of DevS
\citep{Betts2002,Voskuil2003,Wayne2001}. Recent studies reveal that carbon monoxide and 
ascorbic acid environment can also activate this TCS \citep{Kumar2008,Shiloh2008,Taneja2010}.
When phosphorylated at the histidine domain DevS transfers its phosphate group to the aspertate
domain of DevR. The phosphorylated DevR ($R_p$) acts as transcription factor for $\sim$48
genes as well as exerts positive feedback on its own operon. Most of the genes controlled by
$R_p$ contain 20 bp long palindromic sequence (the Dev box) in their upstream region where 
phosphorylated DevR can bind \citep{Park2003}. \textit{Rv3134c} along with \textit{devRS} 
operon contains two such Dev boxes. $R_p$ binds to these two boxes and exerts a strong 
positive feedback as an effect of which \textit{devRS}  is cotranscribed along with \textit{Rv3134c} 
(see Fig.~\ref{network}).

In the present study a theoretical model has been developed that can qualitatively describe 
the dynamical behavior of DevR regulated genes. To this end we have chosen four well studied 
genes Rv3134c, \textit{hspX}, \textit{narK2} and Rv1738 to illustrate DevR controlled regulation 
and effect of different binding sites in the activation of the four genes 
\citep{Chauhan2008a,Chauhan2008b,Chauhan2011}. 
The Rv3134c and \textit{hspX} promoter sites contain two and three DevR binding sites, 
respectively. Whereas, \textit{narK2} and Rv1738 both share the same promoter site 
containing four binding sites (see Fig.~2).
Although these four genes are well studied experimentally further analysis is necessary in 
connection to the complex interaction between DevR and binding sites. Through modeling
we show that how these binding sites control the gene expression individually and collectively. 
In addition, the proposed model simulates temporal dynamics of different mutants that have 
been studied experimentally. From this knowledge we predict temporal dynamics of 
several other mutants which provide qualitative aspects of DevR mediated gene expression.
In addition, we propose a general expression pattern for DevR regulated genes which 
might work well for other DevR controlled gene expression.

We further analyze the proposed model from information theoretical point of view 
to understand the role of different binding sites. Information theory intrinsically takes care 
of generalized concept of communication \citep{Shannon1948}. Information processing
in biological systems has been successfully analyzed using this concept 
\citep{Schneider1990,Schneider1991a,Schneider1991b,Schneider1994,Schneider1997a,Schneider1999,Schneider2000,Hengen1997,Shultzaberger2007}. Using the concept of information
theory we have shown that our model parameters are in linear relationship with the individual 
information of sequences. Another important aspect of information theoretical study is the 
measurement of isothermal molecular efficiency that has a maximum limit of 70\% 
\citep{Schneider2010}. In the present study, we show that DevR controlled promoter
sequences have efficiency around 60-65\%, thus following the general trend of isothermal 
efficiency.

%%%%% Figure 1

%\begin{figure}[!t]
%\includegraphics[width=0.75\linewidth,angle=0]{devRS-network.eps}
%\includegraphics[width=0.75\linewidth,angle=0]{devRS-network.pdf}
%\caption{(color online) Schematic diagram of the signal transduction network in DevR-DevS 
%two component system. The positive feedback of DevR on its own operon and on
%Rv3134c is shown by the dotted line from phosphorylated DevR to the two binding sites 
%S (distal) and P (proximal) denoted by open boxes. $T_H$ denotes hypoxia inducible 
%promoter for Rv3134c. For simplicity we have omitted mRNA and degradation of proteins 
%in the diagram.
%}
%\label{network}
%\end{figure}

%%%%%%%%%% The Model

\section{The model}

To understand the dynamics of DevR regulated genes in \textit{M. tuberculosis}, we 
propose in the following a theoretical model based on mass action kinetics of 
DevR-promoter interaction. The proposed model describes qualitative features of the 
wild type strain as well as the behavior of some novel mutants. Objectives of the 
proposed model are following. First, the developed model has been utilized to 
describe temporal dynamics of DevR regulatory genes in terms of fold induction. 
Second, after being successful in reproducing qualitative features of the wild type 
strain, we make \textit{in silico} testable predictions for some novel mutants.

%%%%%%%%%% The Operon

\subsection{{\rm Rv3134c}-devRS operon}

As mentioned earlier a typical TCS consists of a periplasmic sensor domain and a 
cytoplasmic response regulator and most importantly this composite system needs 
a stimulus to make the circuit operative. Similar to other TCS, DevRS gets activated 
under hypoxic condition \citep{Park2003}. Here DevS and DevR are the sensor protein 
and the response regulator protein, respectively. Once the system is active, DevS gets 
auto-phosphorylated at the histidine residue and forms phosphorylated DevS which 
then transfers the phosphate group to its cognate partner DevR to generate pool of 
phosphorylated DevR. Phosphorylated response regulator ($R_p$) binds to two 
upstream binding sites of its own operon that leads to co-transcription of Rv3134c 
along with \textit{devRS} (see Fig.~\ref{network}). Dual binding at the promoter site 
is necessary, as mutation (single or double) at these binding sites causes loss in 
gene expression \citep{Chauhan2008a}.

Outcome of the activation of Rv3134c-\textit{devRS} is to generate pool of phosphorylated 
response regulator ($R_p$) that controls several downstream genes. In the present model,
we follow a simple mechanism for generation of $R_p$
\begin{equation}
\label{pool}
\overset{k_{srp}}{\longrightarrow} Rp \overset{k_{drp}}\longrightarrow \varnothing .
\end{equation}

\noindent
Eq.~(\ref{pool}) takes care of generation and removal of the pool of phosphorylated DevR
that acts as a transcription factor for the downstream genes. Since in the present study we 
are interested only in the dynamics of DevR regulated genes, the minimal kinetics for the
generation of $R_p$ is sufficient to study the dynamics of the downstream genes.

%%%%% Figure 2

%\begin{figure}[!t]
%\begin{center}
%\includegraphics[width=0.75\linewidth,angle=0]{devrs-fig2.eps}
%\includegraphics[width=0.75\linewidth,angle=0]{devrs-fig2}
%\end{center}
%\caption{(color online) Schematic diagram of phosphorylated DevR interaction with different 
%binding sites (open boxes) of Rv3134c, \textit{hspX}, \textit{narK2} and Rv1738. Rv3134c and 
%\textit{hspX} contains two and three binding sites, respectively. Both \textit{narK2} and Rv1738 
%share same promoter site containing four binding sites.
%}
%\label{genes}
%\end{figure}

%%%%%%%%%% Downstream Genes

\subsection{DevR regulated genes}

\begin{figure}[!t]
\begin{center}
\includegraphics[width=0.75\linewidth,angle=0]{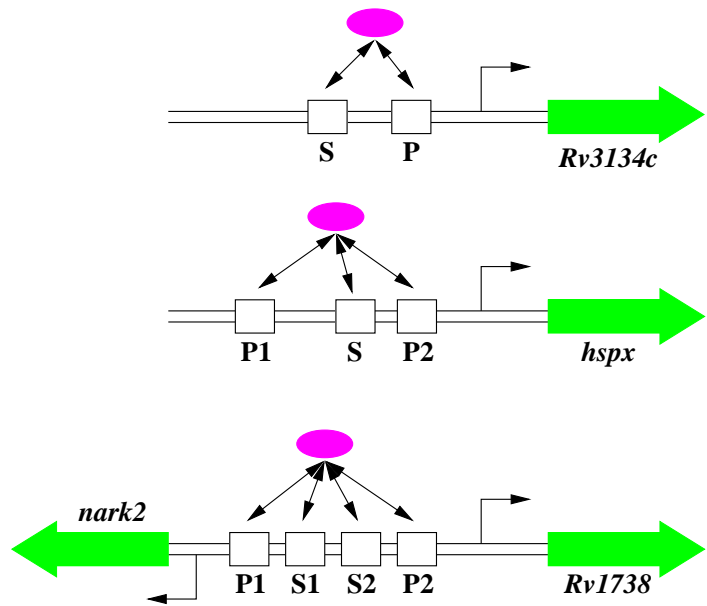}
\end{center}
\caption{Schematic diagram of interaction of phosphorylated DevR with different 
binding sites (open boxes) of Rv3134c, \textit{hspX}, \textit{narK2} and Rv1738. 
Rv3134c and \textit{hspX} contains two and three binding sites, respectively. 
\textit{narK2} and Rv1738 share same promoter containing four binding sites.
}
\label{genes}
\end{figure}

Under hypoxic condition DevRS TCS regulates $\sim$48 genes which are broadly 
classified into four classes, according to the number of Dev boxes (DevR binding site) 
present in the promoter site \citep{Chauhan2011}. 
In the present work we only deal with four genes (Rv3134c,  \textit{hspX}, \textit{narK2} 
and Rv1738) which have been extensively studied experimentally \citep{Chauhan2008b}. 
The main reason behind choosing these genes is that they are well characterized and 
experimental data for mutation in the binding sites of these genes are available 
\citep{Chauhan2008b}. In addition, experimental temporal dynamics of these genes 
serves as an excellent basis for validation of our theoretical model. Before proceeding 
further, we would like to mention that the four selected genes can be categorized into 
three different classes based on the number of available DevR binding boxes 
(see Fig.~\ref{genes}). The most simple case is Rv3134c containing only two Dev boxes 
and belongs to class I. In \textit{hspx}, that belongs to class II, there are three Dev boxes. 
A complex four DevR binding box structure is present for genes \textit{nark2}-Rv1738 and 
are grouped into class III.

%%%%%%%%%% Rv3134c

\subsubsection{{\rm Rv3134c}}

Rv3134c gene is the simplest in construct compared to other DevR regulated genes. 
It has two Dev boxes, one is primary and another is secondary. Phosphorylated DevR 
($R_p$) binds to both the primary ($P$) and the secondary ($S$) binding sites 
(Fig.~\ref{genes}). Kinetics for binding of $R_p$ to these sites can be modeled as
\begin{eqnarray}
P+R_p &\overset{k_{b1}}{\underset{k_{u1}}{\rightleftharpoons}} & P^\ast , \\
S+R_p &\overset{k_{b2}}{\underset{k_{u2}}{\rightleftharpoons}} & S^\ast .
\end{eqnarray}

\noindent In the above two equations P, S and P$^\ast$ , S$^\ast$ stand for inactive and 
active states of the primary and the secondary binding sites, respectively. Under non-inducing 
condition both primary and secondary binding sites do not produce any basal level of 
mRNA whereas the activated sites transcribe in a bulk amount,
\begin{eqnarray}
P^\ast \overset{k_{sm1}}{\longrightarrow} mGFP_{4c} ,\\
S^\ast \overset{k_{sm2}}{\longrightarrow} mGFP_{4c} ,\\
P^\ast S^\ast \overset{k_{sm3}}{\longrightarrow} mGFP_{4c} .
\end{eqnarray}

\noindent Here $mGFP_{4c}$ is the mRNA transcribed from Rv3134c. The rate 
constants $k_{sm1}$ and $k_{sm2}$ give a measure of individual contribution from 
primary and secondary binding sites, respectively, and $k_{sm3}$ is the measure of 
co-operative contribution to the transcription of mRNA. The logic behind assuming this 
kind of equations is the following. When transcription factor binds to any single site 
(primary or secondary) it is ready to generate transcripts. If both of the sites are occupied 
by transcription factor  then one helps another and a co-operative effect comes into play 
to produce large amount of transcripts, compared to the transcripts generated from single 
site occupancy (primary or secondary) as observed in the wild type and single binding 
box deleted mutants \citep{Chauhan2008a}. Advantage of this approach is that, if only one 
site is occupied, the cooperative contribution becomes zero automatically, which helps us 
to generate temporal dynamics of different mutants.

In addition to the above transcription kinetics we consider natural degradation of produced 
mRNA
\begin{equation}
mGFP_{4c} \overset{k_{dm}}\longrightarrow \varnothing.
\end{equation}

\noindent The transcribed mRNA then gets translated into protein 
\begin{eqnarray}
mGFP_{4c} & \overset{k_{sg}}\longrightarrow & mGFP_{4c} + GFP, \\
GFP & \overset{k_{dg}}\longrightarrow & \varnothing,
\end{eqnarray}

\noindent where $GFP$ is the translated protein with a natural degradation given by Eq.~(9).
It is important to mention that in the experimental setup a promoter-GFP construct has been
used to study the promoter activity \citep{Chauhan2008a,Chauhan2008b}. This we
incorporate in the present model through production of GFP out of the transcripts 
generated from the promoter. For the rest of the promoters (\textit{hspX}, \textit{narK2} and 
Rv1738), we follow the same strategy to study \textit{in silico} promoter activity.

Rv3134c is the operon for DevR regulon and has two binding sites, one is primary and 
another is secondary. It is not really clear whether the $S$ binding site which we are 
considering as \textit{secondary} is actually secondary or not because if one observes 
$P$ and $S$ sites closely there is virtually no difference. From information theoretical 
analysis \citep{Chauhan2011} it is also evident that both sites have almose the same 
Ri value (18.2 for $P$ and 18.3 for $S$), which means almost same amount of energy 
dissipation occurs during binding. Moreover, according to the sequence walker method 
\citep{Schneider1997b} both sites have almost identical contact with protein during binding.
This suggest that architecture of this promoter could be $P$-$P$ rather than $P$-$S$. Since, 
both sites are almost identical in structure, one would expect equal contribution from both 
of them in transcription and hence we assume individual contribution of both binding sites 
to be equal.

%%%%%%%%%% hspx

\subsubsection{hspX}

Promoter of \textit{hspX} gene contains three binding sites of which two are primary
and one is secondary \citep{Chauhan2008b}. Out of the two primary binding sites one is 
proximal to the transcription start point while the other is distal. While modeling dynamics
of these binding sites we have denoted the proximal primary binding site as $P2$, the
distal primary binding site as $P1$ and the secondary binding site as $S$. When 
transcription factor binds to these sites they become active,
\begin{eqnarray}
P1+R_p &\overset{k_{b3}}{\underset{k_{u3}}{\rightleftharpoons}} & P1^\ast , \\
P2+R_p &\overset{k_{b4}}{\underset{k_{u4}}{\rightleftharpoons}} & P2^\ast ,  \\
S+R_p &\overset{k_{b5}}{\underset{k_{u5}}{\rightleftharpoons}} & S^\ast .
\end{eqnarray}

\noindent The activated states P1$^\ast$, P2$^\ast$ and S$^\ast$ are ready for making the 
transcripts. As before, we take individual contribution as well as collective effect during 
production of transcripts
\begin{eqnarray}
P1^\ast & \overset{k_{sm4}}{\longrightarrow} & mGFP_{px} , \\
P2^\ast & \overset{k_{sm5}}{\longrightarrow} & mGFP_{px} , \\
S^\ast    & \overset{k_{sm6}}{\longrightarrow}  & mGFP_{px} , \\
P1^\ast P2^\ast S^\ast & \overset{k_{sm7}}{\longrightarrow} & mGFP_{px} , \\
mGFP_{px} & \overset{k_{dm}}\longrightarrow & \varnothing .
\end{eqnarray}

\noindent From the generated transcripts $mGFP_{px}$ we have considered synthesis 
of proteins along with its degradation,
\begin{eqnarray}
mGFP_{px} & \overset{k_{sg}}\longrightarrow & mGFP_{px} + GFP , \\
GFP & \overset{k_{dg}} \longrightarrow & \varnothing .
\end{eqnarray}

\noindent In \textit{hspx} as $P2$ and $S$ are nearer to the transcription start point they 
mostly control expression of this gene, which can be verified by observing the individual 
contribution of these two sites in the model. As the distance between $P1$ and $S$ is large, 
$P1$ hardly helps to incorporate co-operative effect and hence plays little role in regulating 
expression of \textit{hspX} which we will discuss later.

%%%%%%%%%% narK2-Rv1738

\subsubsection{narK2-{\rm Rv1738}}

This system has four binding sites among which two are primary ($P1$ and $P2$) and two are
secondary ($S1$ and $S2$). $P1$ and $S1$ are nearer to the transcription start site of 
\textit{narK2} compared to $P2$ and $S2$. Note that the binding sites $P1$, $S2$ and $P2$ 
have been identified earlier and was denoted as $D1$, $D2$ and $D3$, respectively  
\citep{Chauhan2008b}. The binding site $S1$ was identified later \citep{Chauhan2011}. 
While developing our model we have followed the recent nomenclature \citep{Chauhan2011}. 
Earlier we have mentioned that the proximal 
binding sites play a major role in transcription compared to the distal binding sites. So $P1$ 
and $S1$ contribute mostly to the transcription of \textit{narK2} not only because they are 
located nearer but also for the co-operative effect between them. Similarly, for the Rv1738 
promoter, $P2$ and $S2$ are nearer to the transcription start site and hence contribute more 
than $P1$ and $S1$ towards making the transcripts. In addition, there is also a cooperative 
effect between them. As the distance between $P1$, $S2$ and $P2$, $S1$ is large, collective 
cooperative effect can not be operative here. There are always a relative competetion between 
two pairs as they are transcribing in opposite direction and share the same promoter site. 
Experimental result suggest that expression of Rv1738 remains always high compared to
expression of \textit{narK2} \citep{Chauhan2008b}.
At this point it is important to mention that in the \textit{narK2}-Rv1738 system transcriptional
interference is operative due to overlapping divergent promoter structure \citep{Shearwin2005}.
However, to keep the model simple we do not consider the mechanism of transcriptional
interference in the present work.

Similar to the previous cases we first generate the activated states of each binding sites 
as follows,
\begin{eqnarray}
P1+R_p &\overset{k_{b6}}{\underset{k_{u6}}{\rightleftharpoons}} & P1^\ast , \\
P2+R_p &\overset{k_{b7}}{\underset{k_{u7}}{\rightleftharpoons}} & P2^\ast , \\
S1+R_p &\overset{k_{b8}}{\underset{k_{u8}}{\rightleftharpoons}} & S1^\ast , \\
S2+R_p &\overset{k_{b9}}{\underset{k_{u9}}{\rightleftharpoons}} & S2^\ast .
\end{eqnarray}

\noindent The activated states are able to generate transcripts for both 
\textit{narK2} and Rv1738,
\begin{eqnarray}
P1^\ast \overset{k_{sm8}}{\longrightarrow} mGFP_{K2} , \\
P1^\ast \overset{k_{sm9}}{\longrightarrow} mGFP_{38} , \\
P2^\ast \overset{k_{sm10}}{\longrightarrow} mGFP_{K2} , \\
P2^\ast \overset{k_{sm11}}{\longrightarrow} mGFP_{38} , \\
S1^\ast \overset{k_{sm12}}{\longrightarrow} mGFP_{K2} , \\
S1^\ast \overset{k_{sm13}}{\longrightarrow} mGFP_{38} , \\
S2^\ast \overset{k_{sm14}}{\longrightarrow} mGFP_{K2} ,  \\
S2^\ast \overset{k_{sm15}}{\longrightarrow} mGFP_{38} , \\
P1^\ast S1^\ast \overset{k_{sm16}}{\longrightarrow} mGFP_{K2} , \\
P1^\ast S1^\ast \overset{k_{sm17}}{\longrightarrow} mGFP_{38} , \\
P2^\ast S2^\ast \overset{k_{sm18}}{\longrightarrow} mGFP_{K2} , \\
P2^\ast S2^\ast \overset{k_{sm19}}{\longrightarrow} mGFP_{38} .
\end{eqnarray}

\noindent Degradation of mRNA and GFP production have been modeled in a
similar fashion,
\begin{eqnarray}
mGFP_{K2}  & \overset{k_{dm}}\longrightarrow & \varnothing , \\
mGFP_{38}  & \overset{k_{dm}}\longrightarrow & \varnothing , \\
mGFP_{K2}  & \overset{k_{sg}} \longrightarrow & mGFP_{K2} +GFP , \\
mGFP_{38}  & \overset{k_{sg}} \longrightarrow & mGFP_{38} + GFP , \\
GFP & \overset{k_{dg}}\longrightarrow & \varnothing .
\end{eqnarray}

\noindent As mentioned in the work of \cite{Chauhan2011}, for \textit{narK2} promoter 
$P1$ and $S1$ play a major role whereas for Rv1738 promoter $P2$ and $S2$ play 
the major role, though all of them are common for both promoters. This in turn affects 
behavior of their mutants. So, there is a clear division among these four binding sites 
and moreover, secondary binding sites basically help the primary binding sites through 
co-operative effect. Hence, while formulating our model we have taken two co-operative 
contributions. By choosing proper parameter values of the rate constants for the above 
kinetics, one can describe temporal dynamics of wild type and various mutants in 
terms of fold induction of GFP, which we have discussed in the next section.

From the above discussion, one can generalise the model and can describe dynamics of 
any DevR regulated promoter. For a promoter site containing $N$ number of binding sites
the kinetics will be,
\begin{eqnarray}
P_i+R_p \overset{k_{bi}}{\underset{k_{ui}}{\rightleftharpoons}}  P_i^\ast , \\
P_i^\ast  \overset{k_{smi}}{\longrightarrow}  mGFP 
{\overset{k_{d,m}}\longrightarrow}  \varnothing , \\
{P_i^\ast \ldots P_N^\ast} \overset{k_{smc}}{\longrightarrow} mGFP
{\overset{k_{d,m}}\longrightarrow} \varnothing , \\
mGFP \overset{k_{sg}}\longrightarrow mGFP + GFP  \\
GFP \overset{k_{dg}}\longrightarrow \varnothing .
\end{eqnarray}
Where $k_{smi}$ ($i \in \{1, N \}$) are the individual contribution of the $N$-th binding site 
and $k_{smc}$ are the measure of co-operative contribution from all the $n$ binding sites.

%%%%%%%%%% Results and Discussions

\section{Results and Discussions}

To check the validity of our proposed model, developed in the previous section, the kinetic 
equations (1-40) have been translated into sets of nonlinear ordinary differential equations 
(ODEs) (see Appendix). To study temporal behavior of the wild type strain and the different 
mutants, sets of nonlinear ODEs are solved by XPP (http://www.math.pitt.edu/~bard/xpp/xpp.html) 
using the parameter set given in Tables~1-2. The parameter set listed in the tables were 
guessed to generate the temporal experimental profile given in Figures~3,9,10.

\subsection{Wild type}

\begin{figure}[!t]
\begin{center}
\includegraphics[width=0.75\linewidth,angle=0]{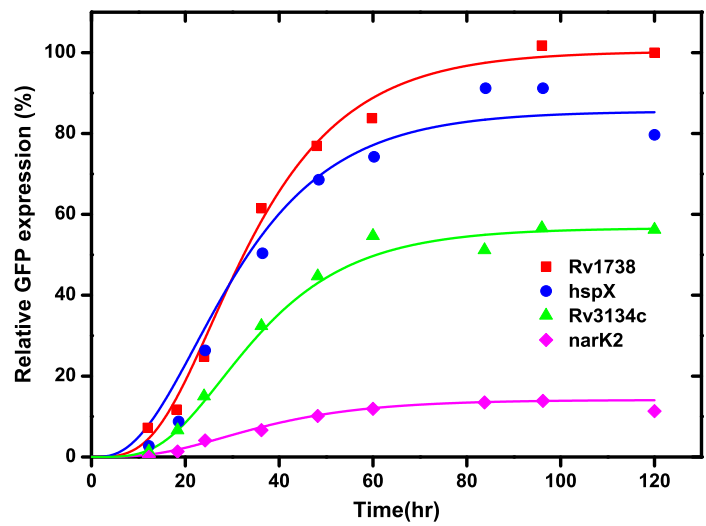}
\end{center}
\caption{Time evolution of relative GFP expression of Rv3134c and three
downstream genes. Symbols are taken from \cite{Chauhan2008b} 
and continuous lines are results of numerical simulation.
}
\label{allwt}
\end{figure}

In Fig.~\ref{allwt}, we compare numerical results with experimental data 
for time evolution of relative GFP level for the promoters of
Rv3134c, \textit{hspx}, \textit{narK2} and Rv1738. In the work of \cite{Chauhan2008b}
GFP levels have been measured in the unit of RFU/OD. To compare the experimental 
data with numerical simulation results, we have scaled all the experimental data by the 
maximum expression level of Rv1738, (among the four genes Rv1738 is the most expressive 
one; see Fig.~8 of \cite{Chauhan2008b}) so that
we get a dimensionless relative expression level for all the four genes. For comparison
numerical data has been scaled by using the same strategy. From Fig.~\ref{allwt} it is evident 
that our model captures the qualitative aspects of the \textit{in vivo} experimental results. 
In addition, our model could reproduce the competition between \textit{narK2} and Rv1738 
as they share same promoter.

%%%%% Figure 3

%\begin{figure}[!t]
%\begin{center}
%\includegraphics[width=0.75\linewidth,angle=0]{all-wild-type.eps}
%\includegraphics[width=0.75\linewidth,angle=0]{all-wild-type.pdf}
%\end{center}
%\caption{(color online) Time evolution of relative GFP expression of Rv3134c and three
%downstream genes. Symbols are taken from \cite{Chauhan2008b} and continuous lines are the 
%results of numerical simulation.
%}
%\label{allwt}
%\end{figure}

%%%%% Figure 4

%\begin{figure}[!t]
%\begin{center}
%\includegraphics[width=0.75\linewidth,angle=0]{3134c-mutants-cartoon.eps}
%\includegraphics[width=0.75\linewidth,angle=0]{3134c-mutants-cartoon.pdf}
%\end{center}
%\caption{(color online) Possible mutants by permutation of two binding sites of Rv3134c 
%promoter region. All the three mutants have been studied by \cite{Chauhan2008a}.
%}
%\label{3134c}
%\end{figure}

%%%%% Figure 5

%\begin{figure}[!b]
%\begin{center}
%\includegraphics[width=0.75\linewidth,angle=0]{3134c-mutants.eps}
%\includegraphics[width=0.75\linewidth,angle=0]{3134c-mutants.pdf}
%\end{center}
%\caption{(color online) Time evolution of the wild type and both the mutants pmutP and pmutS for
%Rv3134c have been shown in this figure. The expressions of mutants are significantly low which 
%is shown by axis breaking. For the mutant pmutPS the expression vanishes completely, hence 
%is not shown in the figure.
%}
%\label{mut3134c}
%\end{figure}

\subsection{The mutants}

Being successful in describing temporal evolution of the four DevR regulated wild type 
strains we now look at behavior of their respective mutants. While generating behavior
of a specific mutant we have set the value of respective binding and unbinding rate 
constant to zero.

\begin{figure}[!t]
\begin{center}
\includegraphics[width=0.75\linewidth,angle=0]{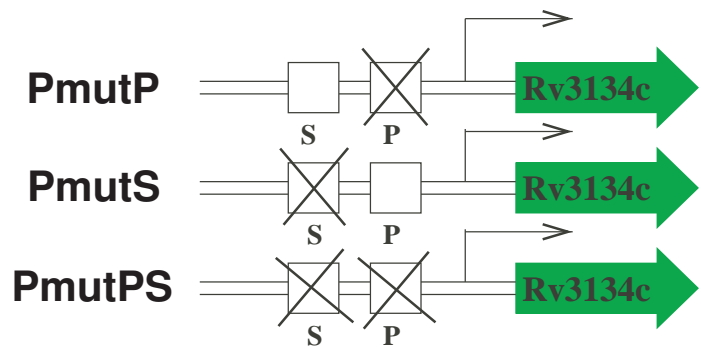}
\end{center}
\caption{Possible mutants by permutation of two binding sites of Rv3134c 
promoter region. All the three mutants have been studied by 
\cite{Chauhan2008a}.
}
\label{3134c}
\end{figure}

\begin{figure}[!b]
\begin{center}
\includegraphics[width=0.75\linewidth,angle=0]{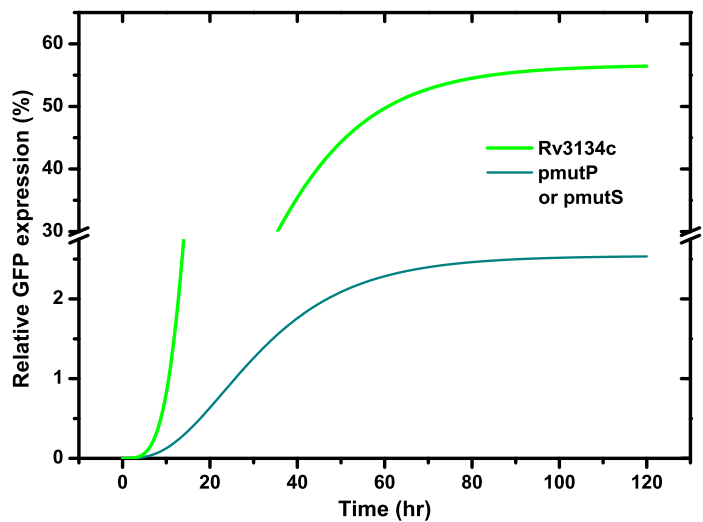}
\end{center}
\caption{Time evolution of Rv3134c (wild type and mutants (pmutP and pmutS)). 
Expression of mutants is significantly low which is shown by axis breaking. According 
to our model, expression of the double mutant pmutPS vanishes completely, hence 
is not shown in the figure. 
}
\label{mut3134c}
\end{figure}

As mentioned earlier Rv3134c promoter has two binding sites, one is primary and other is 
secondary. However, from their interaction with DevR it is not easy to detect the difference.
Only way to make the distinction is to look at the expression level of different mutants
(see Fig.~\ref{3134c}).
When the primary site is mutated (pmutP) the expression level decreases as expected 
but on the other hand, expression level is very much similar compared to pmutS where 
secondary site has been mutated. But for pmutPS, where both sites are mutated,
expression level is also same as pmutP or pmutS. This is quite unlikely as both of them should 
contribute to the generation of transcript. It might happen that, other than $P$ and $S$ there 
is a third binding site that contributes to the expression of pmutPS strain. This needs further 
careful experimental verification. It is important to note that,  expression level of the double 
mutant pmutPS is almost not detectable from our model. According to \cite{Chauhan2008a},
expression of pmutP or pmutS decreases $\sim$25 fold compared to the wild type expression 
at 48 hours which is very much close to our simulation result (see Fig.~\ref{mut3134c}).

%%%%% Figure 6

%\begin{figure}[!t]
%\begin{center}
%\includegraphics[width=0.75\linewidth,angle=0]{hspx-mutants-cartoon.eps}
%\includegraphics[width=0.75\linewidth,angle=0]{hspx-mutants-cartoon.pdf}
%\end{center}
%\caption{(color online) Possible mutants by permutation of three binding sites of 
%\textit{hspX} promoter region. The first and the third mutant from top have been studied by 
%\cite{Park2003} and the behavior of other mutants have been predicted in this study.
%}
%\label{hspx}
%\end{figure}

\begin{figure}[!t]
\begin{center}
\includegraphics[width=0.75\linewidth,angle=0]{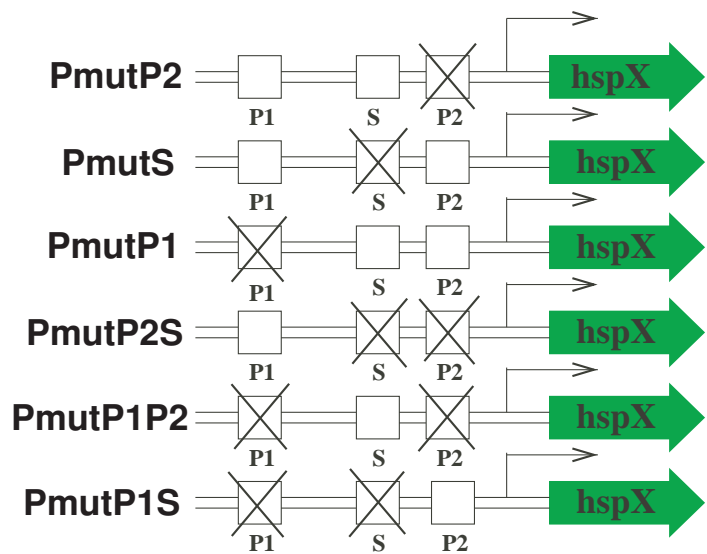}
\end{center}
\caption{Possible mutants by permutation of three binding sites of 
\textit{hspX} promoter region. The first and the third mutant from top have been studied by 
\cite{Park2003} and behavior of other mutants has been predicted in this 
study.
}
\label{hspx}
\end{figure}

\begin{figure}[!b]
\begin{center}
\includegraphics[width=0.75\linewidth,angle=0]{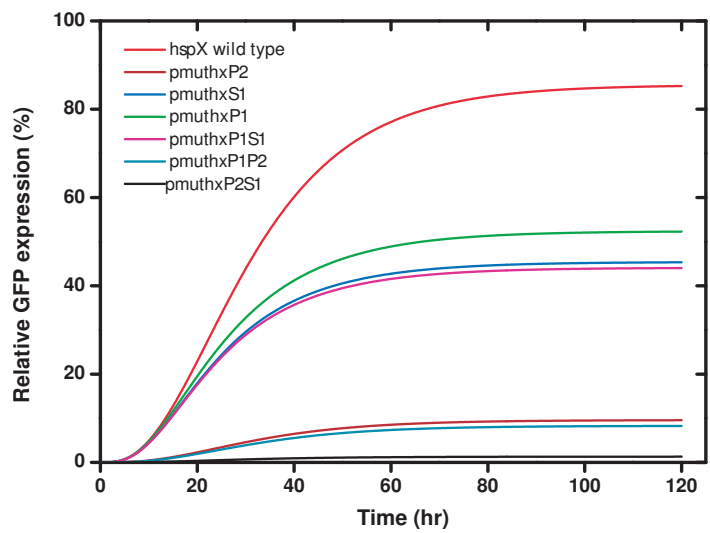}
\end{center}
\caption{Time evolution of wild type \textit{hspX} and all its mutants. All the 
double mutants except pmutP2S1 and pmutP1 have very low expression 
showing the importance of $P1$ binding site.
}
\label{muthspx}
\end{figure}

\textit{hspX} promoter has three binding sites, two are primary and one is secondary 
(see Fig.~\ref{hspx}). Among the two primary sites one is proximal, another is distal. 
The two primary binding sites were identified by \cite{Park2003} and the secondary
binding site was identified by \cite{Chauhan2011}. When the distal primary site $P1$ 
is mutated it can recover GFP expression level $\sim 70\%$ of wild type expression but 
if the proximal binding site $P2$ is mutated it recovers $\sim 53\%$ of wild type expression 
which is also revealed from our model (see Fig.~\ref{muthspx}). Interestingly, when mutations 
are done on both primary sites ($P1$ and $P2$) our model shows a minimal expression 
($\sim 12\%$ of wild type expression) which is in agreement with the experimental data of 
\cite{Park2003}. We have also created double mutants pmutP2S and pmutP1S \emph{in silico} 
where expression level for pmutP2S1 is undetectable and shows the importance of nearer 
binding site ($P2$ and $S$) on gene expression.

%%%%% Figure  7

%\begin{figure}[!b]
%\begin{center}
%\includegraphics[width=0.75\linewidth,angle=0]{hspx-mutants.eps}
%\includegraphics[width=0.75\linewidth,angle=0]{hspx-mutants.pdf}
%\end{center}
%\caption{(color online) Time evolution of wild type \textit{hspX} and all its mutants. All the 
%double mutants except pmutP2S1 and pmutP1 have very low expression that shows
%the importance of $P1$ binding site.
%}
%\label{muthspx}
%\end{figure}

%%%%% Figure 8

%\begin{figure}[!t]
%\begin{center}
%\includegraphics[width=0.85\linewidth,angle=0]{nark2-rv1738-mutants-cartoon.eps}
%\includegraphics[width=0.85\linewidth,angle=0]{nark2-rv1738-mutants-cartoon.pdf}
%\end{center}
%\caption{(color online) Possible mutants by permutation of four binding sites of 
%\textit{narK2}-Rv1738 intergenic promoter region. The first three and fifth mutants from top
%have been created by \cite{Chauhan2008b} and the behavior of other mutants have been 
%predicted in this study.
%}
%\label{nark2-rv1738}
%\end{figure}

Among the four binding sites in the intergenic sequence of \textit{narK2}-Rv1738 promoter 
$P1$ and $S1$ majorly control the transcription of \textit{narK2}. On the other hand,
$P2$ and $S2$ control the transcriton of Rv1738. Effect of Dev box mutation for this 
promoter has been studied by \cite{Chauhan2008b} by means of GFP reporter assay. 
At this point it is important to mention that, two sets of mutant data are available in the
literature, one is for plate format and the other is for tube format. The main difference 
between the two format is duration of experiment. The tube format needed twenty days 
for complete monitoring of the assay and the plate format needed five days. As the tube 
format takes larger time, there might be food limitations and other factors affecting growth 
of the colony and hence nonlinear degradation of proteins may play a role during the 
experiment. As we do not explicitly incorporate nonlinear degradation of proteins in our 
model, we only consider experimental data obtained from the plate format.

Out of the four binding sites present in the intergenic region of Rv1738-\textit{narK2}, 
$S1$ has been identified recently \citep{Chauhan2011} and the other three sites 
($P1$, $P2$ and $S2$) were identified earlier by \cite{Chauhan2008b} 
(see Fig.~\ref{genes}). Secondary binding sites mainly help primary binding sites 
via co-operatively but their contribution alone towards transcription is low.
On the other hand, primary binding sites without any co-operative effect have 
potential to generate a good amount of transcripts. This is also evident from the 
mutational analysis of Dev boxes for \textit{narK2}-Rv1738 system. $G_{4}$, $G_{5}$, 
$G_{6}$ and $C_{8}$ are the most conserved bases for a Dev box which have been 
selectively deleted to create mutants \citep{Chauhan2011}.

%%%%% Figure 9

%\begin{figure}[!t]
%\begin{center}
%\includegraphics[width=0.75\linewidth,angle=0]{nark2-mutants.eps}
%\includegraphics[width=0.75\linewidth,angle=0]{nark2-mutants.pdf}
%\end{center}
%\caption{(color online) Time evolution of relative GFP expression of \textit{narK2} and its mutants.
%Symbols are taken from \cite{Chauhan2008b} and the continuous lines are the results of numerical 
%simulation. According to our model pAmutS2 and pAmutP3 behave equivalently.
%}
%\label{mutnark2}
%\end{figure}

%%%%% Figure 10

%\begin{figure}[!b]
%\begin{center}
%\includegraphics[width=0.75\linewidth,angle=0]{rv1738-mutants.eps}
%\includegraphics[width=0.75\linewidth,angle=0]{rv1738-mutants.pdf}
%\end{center}
%\caption{(color online) Time evolution of relative GFP expression of Rv1738 and its mutants. 
%Symbols are taken from \cite{Chauhan2008b} and the continuous lines are the results 
%of numerical simulation.
%}
%\label{mutrv1738}
%\end{figure}

\begin{figure}[!t]
\begin{center}
\includegraphics[width=0.85\linewidth,angle=0]{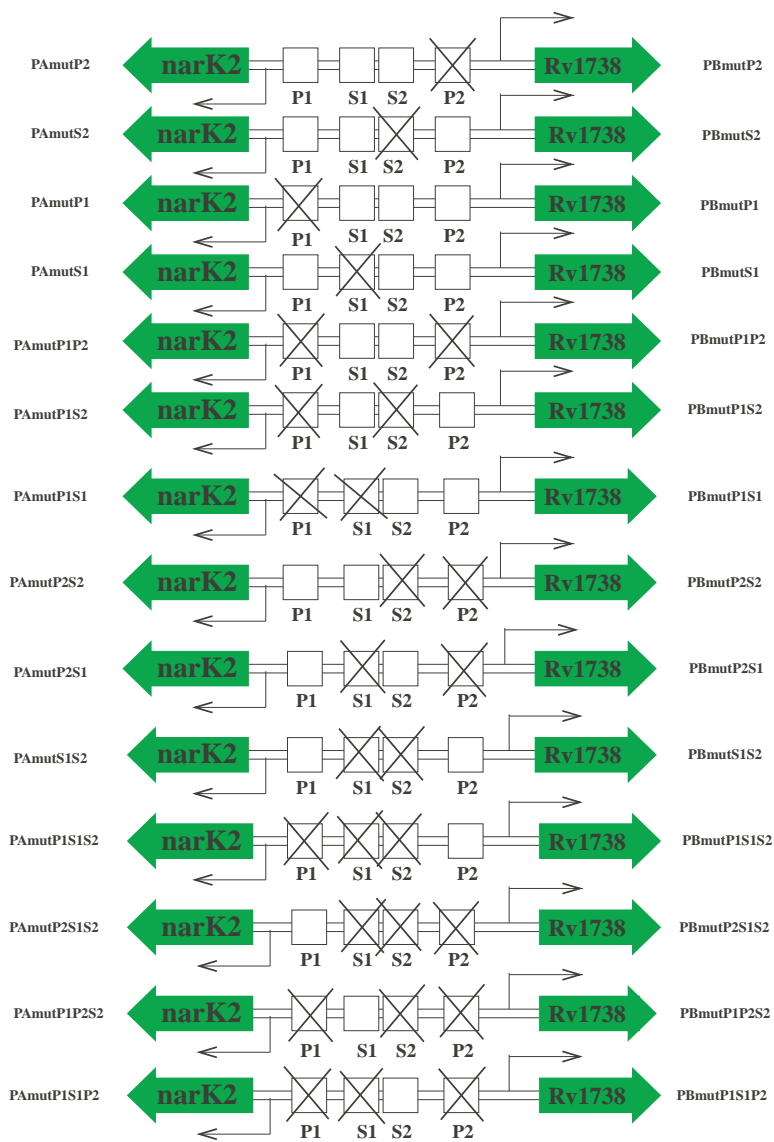}
\end{center}
\caption{Possible mutants by permutation of four binding sites of 
\textit{narK2}-Rv1738 intergenic promoter region. The first three and the fifth mutant
from top have been created by \cite{Chauhan2008b} and behavior of other 
mutants has been predicted in this study.
}
\label{nark2-rv1738}
\end{figure}

\begin{figure}[!t]
\begin{center}
\includegraphics[width=0.75\linewidth,angle=0]{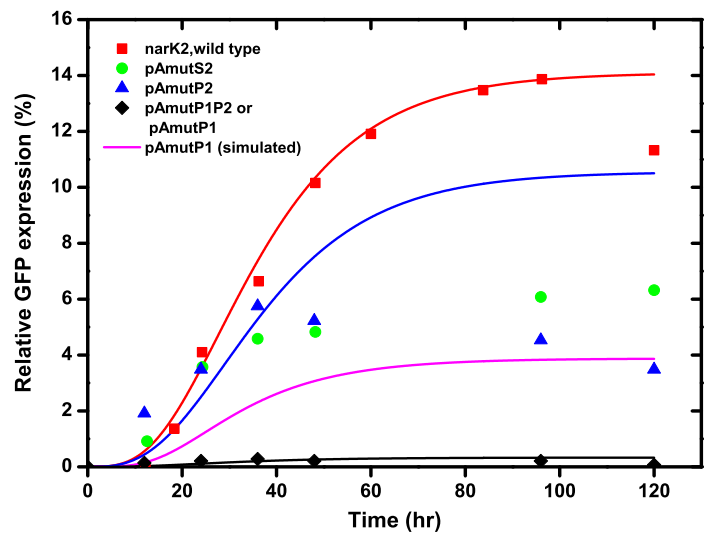}
\end{center}
\caption{Time evolution of relative GFP expression of \textit{narK2} and 
its mutants. Symbols are taken from \cite{Chauhan2008b} and the 
continuous lines are results of numerical simulation. According to our 
model pAmutS2 and pAmutP2 behave equivalently.
}
\label{mutnark2}
\end{figure}

\begin{figure}[!b]
\begin{center}
\includegraphics[width=0.75\linewidth,angle=0]{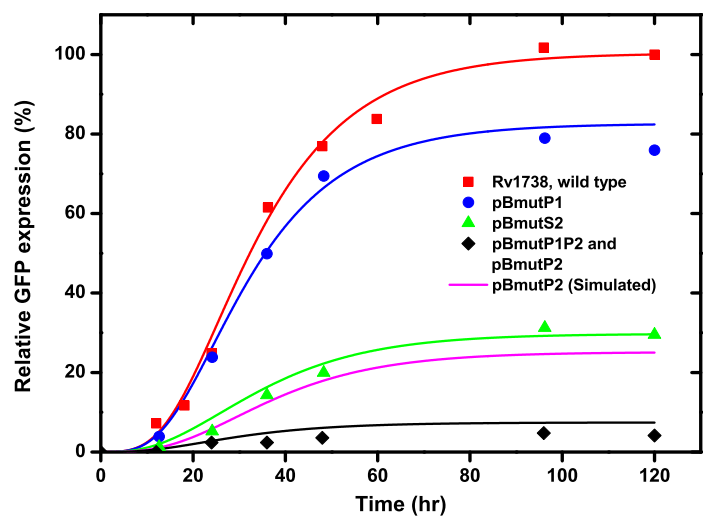}
\end{center}
\caption{Time evolution of relative GFP expression of Rv1738 and its 
mutants. Symbols are taken from \cite{Chauhan2008b} and the 
continuous lines are results of numerical simulation.
}
\label{mutrv1738}
\end{figure}

For the mutant pAmutP1, where mutation has been done on $P1$ (a primary binding site 
for \textit{narK2}) expression level is minimum and is almost not detectable. From this 
result it can be inferred that contributions from $P2$ and $S2$ are really negligible in 
\textit{narK2} expression. In another mutant pBmutP2, $P2$ is mutated. As $P2$ 
mainly controls expression of Rv1738, level of transcript goes down very much 
and is also not detectable. At the same time, expression of \textit{narK2} for this  mutant 
is quite good in comparison to pAmutP1 (see Figs.~\ref{mutnark2},\ref{mutrv1738}). 
Actually, there is no difference in construct between pAmutP2 and pBmutP2, as in both 
case $P2$ site has been mutated (see Fig.~\ref{nark2-rv1738}). For the mutant pBmutP1, 
expression of Rv1738 is 75\% of that of the wild type strain, as $P1$ site is far upstream 
of the transcription start point  of Rv1738 and has little contribution in transcription. 
It is clear from the expression of these mutants that there are two co-operative 
effects operative in \textit{narK2}-Rv1738 system, one between $P1$ and $S1$ and 
another between $P2$ and $S2$ which we have incorporated in our model. From the 
close position of the two secondary sites $S1$ and $S2$ one might expect a third 
co-operative contribution, but surprisingly it does not exist as they together cannot 
recruit $R_p$ to the primary sites. Similarly, when $S2$ is mutated (pAmutS2), we 
observe the same expression level of \textit{narK2} as it was for pAmutP2. But expression 
level decreases for Rv1738 and becomes 25\% of that of the wild type strain. When both 
primary sites are mutated (pAmutP1P2 and pBmutP1P2), as expected, expression level 
for both genes vanishes almost completely.

%%%%% Figure 11

%\begin{figure}[!t]
%\begin{center}
%\includegraphics[width=0.75\linewidth,angle=0]{nark2-prediction-double-mutants.eps}
%\includegraphics[width=0.75\linewidth,angle=0]{nark2-prediction-double-mutants.pdf}
%\end{center}
%\caption{(color online) Prediction for temporal dynamics of relative GFP expression 
%of \textit{narK2} and its double mutants. According to our prediction except pAmutP1S2 
%others should have detectable expression.
%}
%\label{mutnark2dob}
%\end{figure}

%%%%% Figure 12

%\begin{figure}[!b]
%\begin{center}
%\includegraphics[width=0.75\linewidth,angle=0]{nark2-prediction-triple-mutants.eps}
%\includegraphics[width=0.75\linewidth,angle=0]{nark2-prediction-triple-mutants.pdf}
%\end{center}
%\caption{(color online) Prediction for temporal dynamics of relative GFP expression of
%\textit{narK2} and its triple mutants in which only pAmutP2S1S2 should have detectable 
%expression.
%}
%\label{mutnark2trp}
%\end{figure}

\begin{figure}[!t]
\begin{center}
\includegraphics[width=0.75\linewidth,angle=0]{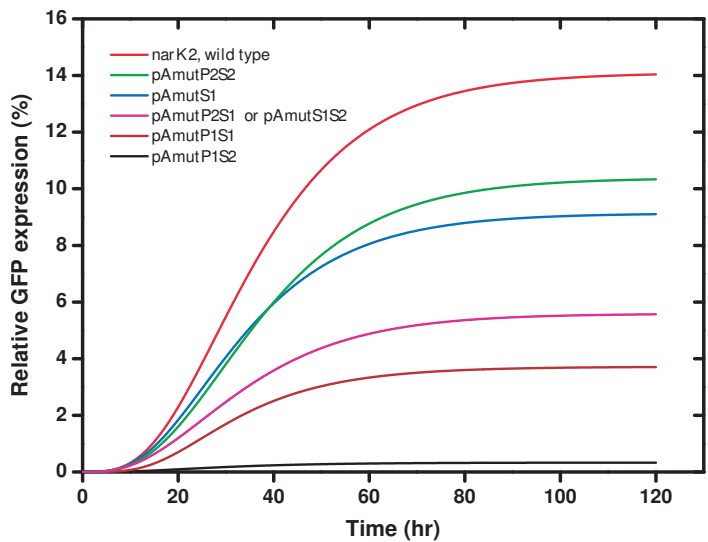}
\end{center}
\caption{Prediction for temporal dynamics of relative GFP expression of \textit{narK2} 
and its double mutants. According to our model, except pAmutP1S2 others should 
have detectable expression.
}
\label{mutnark2dob}
\end{figure}

\begin{figure}[!b]
\begin{center}
\includegraphics[width=0.75\linewidth,angle=0]{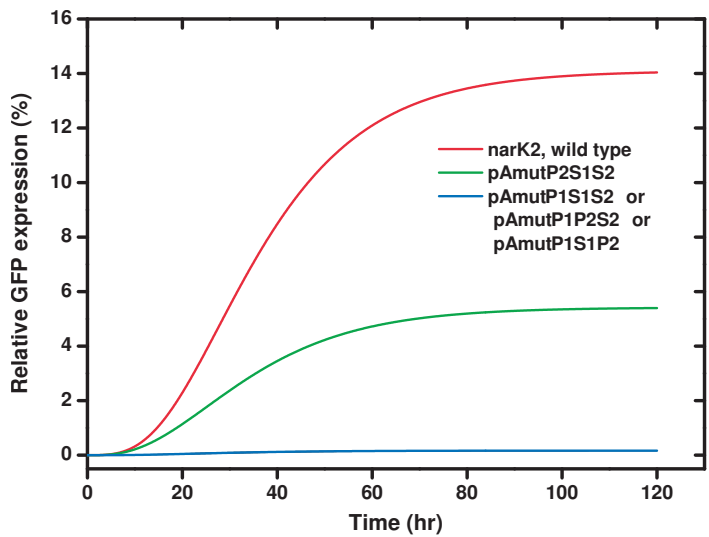}
\end{center}
\caption{Prediction for temporal dynamics of relative GFP expression of
\textit{narK2} and its triple mutants in which only pAmutP2S1S2 should have 
detectable expression.}
\label{mutnark2trp}
\end{figure}

\begin{figure}[!t]
\begin{center}
\includegraphics[width=0.75\linewidth,angle=0]{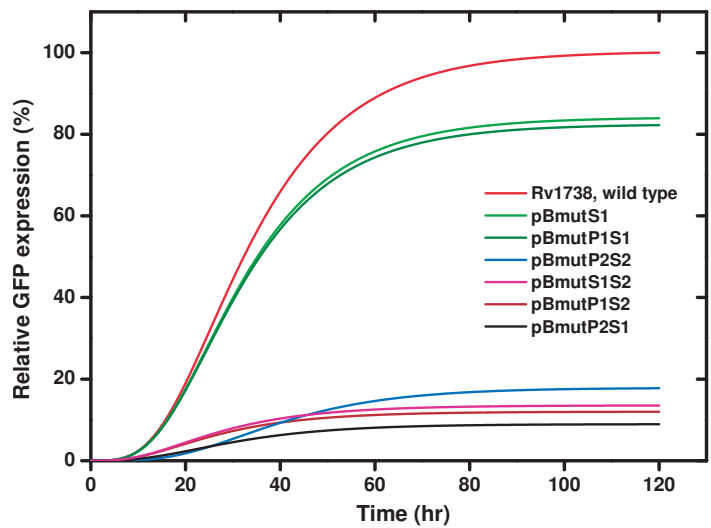}
\end{center}
\caption{Prediction for temporal dynamics of GFP expression of Rv1738 and its 
double mutants. The expression of the double mutants which have either $P2$ or 
$S2$ or both sites mutated are really small. This clarifies the importance of these 
two sites on the expression of Rv1738 gene.}
\label{mutrv1738dob}
\end{figure}

\begin{figure}[!t]
\begin{center}
\includegraphics[width=0.75\linewidth,angle=0]{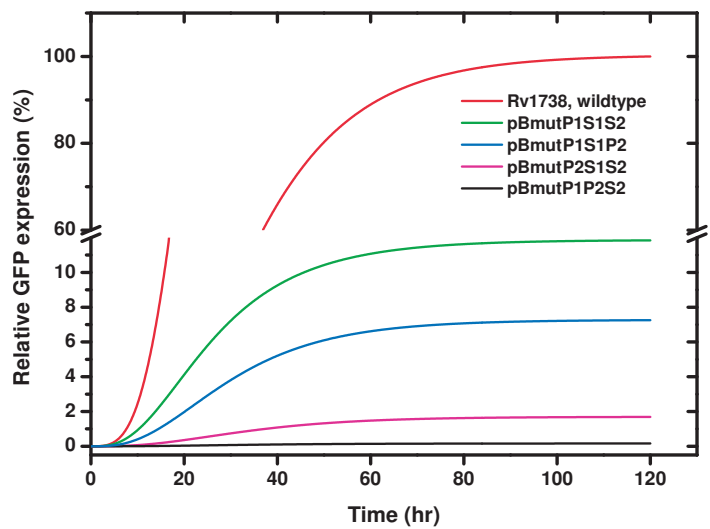}
\end{center}
\caption{Prediction for temporal dynamics of GFP expression of Rv1738 and its 
triple mutants. All the mutants have very low expression comparative to the wild 
type strain, which is shown by the axis break.
}
\label{mutrv1738trp}
\end{figure}

Among the \textit{narK2} double mutants, pAmutP2S1 and pAmutS1S2 have same 
expression, which clearly depicts that contribution from $P2$ and $S2$ are same 
but comparatively lower than $P1$ site, as the expression of pAmutP1S1 and 
pAmutP1S2 are very small compared to the wild type expression 
(see Fig.~\ref{mutnark2dob}). Another interesting point is that expression of 
pAmutP2S2 is nearly 70\% of that of wild type expression but expression of 
pAmutP2S1S2 is only 35\% and pAmutP1P2S2 is almost zero 
(see Figs.~\ref{mutnark2dob},\ref{mutnark2trp}). 
Previously we have mentioned that though individual contribution of secondary 
site is low, co-operative contribution which operates through the secondary site is 
not negligible. This becomes clear from the nature of these mutants. Except 
pAmutP2S1S2, other triple mutants have very low expression due to the very 
obvious reason of $P1$ deletion. By the similar reasoning, except pBmutP1S1, all 
other double mutants of Rv1738 have very low expression compared to the wild type 
strain. This reveals importance of $P2$ and $S2$ on gene expression
(see Fig.~\ref{mutrv1738dob}). The minute difference in the expression between 
pBmutS1 and pBmutP1S1 can be explained due to low contribution of $P1$ in the 
expression of Rv1738 gene, while it plays a vital role for \textit{narK2}. This in fact 
justifies our model which incorporates two cooperative contribution for the 
\textit{narK2}-Rv1738 system. All the triple mutants of Rv1738 have significantly low 
expression due to the loss of dual co-operative contribution (see Fig.~\ref{mutrv1738trp}).

%%%%% Figure 13

%\begin{figure}[!t]
%\begin{center}
%\includegraphics[width=0.75\linewidth,angle=0]{rv1738-prediction-double-mutants.eps}
%\includegraphics[width=0.75\linewidth,angle=0]{rv1738-prediction-double-mutants.pdf}
%\end{center}
%\caption{(color online) Prediction for temporal dynamics of GFP expression of Rv1738 and its 
%double mutants. The expression of the double mutants are really small which have either $P2$ 
%or $S2$ or both sites mutated which clears the importance of these two sites on the expression 
%of Rv1738 gene.  
%}
%\label{mutrv1738dob}
%\end{figure}

%%%%% Figure 14

%\begin{figure}[!b]
%\begin{center}
%\includegraphics[width=0.75\linewidth,angle=0]{rv1738-prediction-triple-mutants.eps}
%\includegraphics[width=0.75\linewidth,angle=0]{rv1738-prediction-triple-mutants.pdf}
%\end{center}
%\caption{(color online) Prediction for temporal dynamics of GFP expression of Rv1738 and its 
%triple mutants. All the mutants have very low expression comparative to the wild type strain 
%which is shown by the axis break.
%}
%\label{mutrv1738trp}
%\end{figure}

From the analysis mentioned above one can conclude that our model is really efficient
in describing the temporal dynamics of wild type strain and some mutants. 
Unfortunately our model could 
not explain the behavior of single mutants of \textit{narK2} gene. If one clearly observes 
the \% expression of \textit{narK2} mutants one can see that they are very low ($\sim$6\%). 
At such a low expression level,
fluctuations in the experimental data play a dominant role, which 
is difficult to ignore even when the experiments are performed in a bulk culture. Probably, 
a stochastic version of the present model will be able to remove the anomaly between 
experimental and theoretical data, which we plan to study in the near future.
However, an important pattern for DevR regulated gene expression that evolves out of
this analysis worth mentioning at this point. If a promoter site has a construct with both 
primary and secondary binding sites with co-operativity in binding between them, then 
mutation in primary  binding site can not be recovered (as revealed from the expression 
level) by the system. But if the same happens with the secondary binding site then the 
system recovers itself partly but not as much as it was in the wild type. Though individual 
contribution of secondary binding site is quite low compared to the primary binding site, 
co-operative effect that comes through the secondary binding site plays an important role 
which can not be ruled out. Though apparently, it may look like the primary binding site 
has major role in transcription, we show that the secondary binding site also play a nontrivial 
role in the gene expression mechanism through co-operativity.

\subsection{Parameter sensitivity analysis}

To check the sensitivity of the parameter set (listed in Tables~1-2) on the steady
state level of mRNA we have used the formalism of total parameter variation 
developed by \cite{Barkai1997}. In this formalism, initially all or a subset of rate 
constants are subjected to random perturbation. In our case, the perturbation is 
drawn from a random gaussian distribution whose mean is the unperturbed value 
of each rate constant. In addition, variance of the random gaussian distribution has 
been considered to be a certain percentage (up to maximum of 10\%) of each rate 
constant. After perturbation, we thus have two sets of parameters. The first set 
consists of unperturbed (reference) rate constants, $k_i^0$ and the second set 
consists of perturbed (modified) rate constants, $k_i$. Using these two parameter 
sets one can compute the level of $mGFP_{38}$, $mGFP_{K2}$, $mGFP_{px}$ 
and $mGFP_{4c}$ at steady state. The sensitivity in four different reference mRNA 
levels with respect to model parameters can be characterized by total parameter 
variation $\log (\kappa)$, where $\log (\kappa) = \sum_{i=1}^N | \log (k_i/k_i^0) |$ 
\citep{Barkai1997}.

\begin{figure}[!t]
\begin{center}
\includegraphics[width=0.75\linewidth,angle=0]{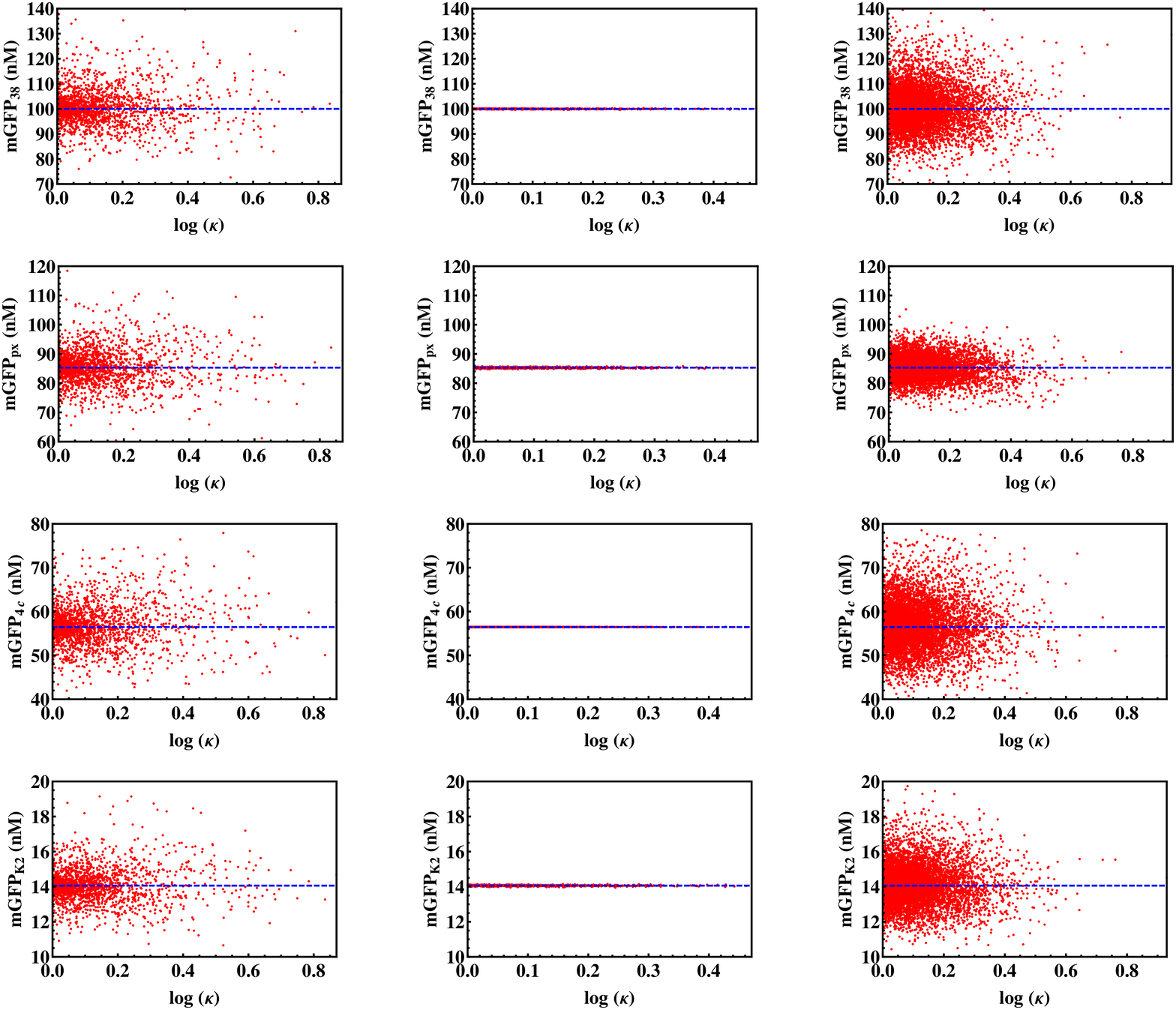}
\end{center}
\caption{Steady state mRNA level as function of total parameter variation 
$\log (\kappa)$. The dashed blue line represents the steady state mRNA
level obtained using the unperturbed parameter set. Each red dot represents 
the same for perturbed parameter set (or subset). 2000 independent simulations 
have been carried out to create the red dots. In the first column all model parameters 
have been perturbed. In the second and third column, parameters related to 
binding-unbinding kinetics and synthesis-degradation kinetics have been 
modified, respectively.
}
\label{sens}
\end{figure}

The results of total parameter variation are shown in Fig.~\ref{sens}. The resultant
data suggests that steady state level of all four mRNAs are sensitive (note the spread
of ordinate) to complete parameter set (first column of Fig.~\ref{sens}) of the model 
which have been modified using the scheme described in the previous paragraph. 
To understand which subset of the complete parameter set is responsible for such 
fluctuations we selectively perturb the parameters related to the binding-unbinding 
kinetics and synthesis-degradation kinetics. When the parameters related to the
binding-unbinding kinetics are perturbed we see that the steady state mRNA level 
are not sensitive to the perturbation (note the collapse of red dots on the dashed blue
line in the second column of Fig.~\ref{sens}). However, when 
the parameter set related to the synthesis-degradation kinetics have been modified we 
see that the modified parameters can recover the fluctuations (third column of Fig.~\ref{sens})
we have observed when the full parameter set has been perturbed (first column of 
Fig.~\ref{sens}). This result suggest that parameters related to synthesis-degradation
kinetics mostly control the steady state mRNA level in our model.

%%%%%%%%%% Information theoretical analysis of devR regulon

\section{Information theoretical analysis of \textit{devR} regulon}

Information theory was founded by Claude E Shannon in late 1940's to analyze the signal
transduction in electrical circuits while addressing the problem of efficient communication 
\citep{Shannon1948,Cover1991}. 
In his work, Shannon showed how to quantify information exchange 
between two electrical devices in the form of bits. Although it is a general theory for electrical 
communication, the basic underlying concept can still be applied to various fields including 
statistical inference, cryptography, quantum computing, networks, communication in 
neurobiology and in recent times in molecular biology too \citep{Borst1999,Rhee2012}.
For example, let us consider the specific or non-specific binding of a transcription factor to
a piece of DNA. It is very much obvious that non-specific binding sites are different from that 
of specific binding sites. One may argue at this point that what makes the binding sites so 
different so that the transcription factor recognizes them differently and binds to these sites so 
specifically and precisely? Another example is the restriction enzyme EcoRI, which cuts the 
pattern 5'-GAATTC-3' throughout the genome. How EcoRI is able to do this so accurately?
These questions can be answered with the help of information theory. The non-specific sites 
do lack of particular information needed for binding of the transcription factor. Exchange of 
information occurs during DNA-protein interaction and hence could be well applicable in the 
present study of DevR mediated transcription of downstream genes.

At this point it is important to connect the mechanism of DNA-protein interaction 
(DevR-promoter interaction considered in the present model) to the
mechanism of a molecular machine. A molecular machine is a macromolecule 
(single or complex) that performs an operation. By operation one means a particular 
task or a specific function that has to be done by the macromolecular machine 
for its survival. For example, when the restriction enzyme EcoRI picks a specific 
sequence (5'-GAATTC-3') from DNA, it acts as a tiny molecular machine capable 
of making decision. According to the theory of molecular machine, binding of EcoRI 
to DNA is restricted (or bounded) by the `machine capacity' while performing the 
specific job \citep{Schneider1991a,Schneider1991b}. This machine capacity is closely 
related to Shannon's `channel capacity'. As long as a molecular machine does not 
exceed its machine capacity, it may perform the task as precise as it needs for survival. 
Following Shannon, the channel capacity can be defined as 
\citep{Shannon1948}
\begin{eqnarray*}
C=W\log_{2}\left(\frac{P}{N}+1\right),
\end{eqnarray*}

\noindent where the bandwidth $W$ defines the range of frequencies used in 
communication and $P/N$ is the `signal to noise' ratio.
In 1959, in the context of satellite communication, efficiency $\epsilon$ has been defined
from the information theoretical point of view \citep{Pierce1959, Raisbeck1963}
\begin{eqnarray*}
\epsilon=\frac{\ln\left(\frac{P}{N}+1\right)}{\frac{P}{N}} .
\end{eqnarray*}

\noindent In 1991 Schneider utilized the above ideas in explaining the mechanism of a
molecular machine \citep{Schneider1991a,Schneider1991b}. Following Schneider's
original notation one can define the channel capacity of a molecular machine as
\begin{eqnarray*}
C_{y}=d_{space}\log_{2}\left(\frac{P_y}{N_y}+1\right),
\end{eqnarray*}

\noindent 
where $d_{space}$ is the number of independent parts of a molecular machine, 
$P_y$ is the energy dissipated per operation and $N_y$ is the thermal noise that 
interferes with the machine during operation. By dividing $P_y$ by the machine 
capacity $C_y$, one gets to know the number of joules that must be dissipated to 
gain one bit of information,
\begin{eqnarray*}
\epsilon \equiv \frac{P_{y}}{C_{y}} \quad (\text{joules per bit}) .
\end{eqnarray*}

\noindent The minimum energy dissipation can be calculated from the channel capacity 
or using the second law of thermodynamics as follows
\begin{eqnarray*}
\epsilon_{min} =k_B T\ln(2) \quad (\text{joules per bit}) ,
\end{eqnarray*}

\noindent where $k_B$ is the Boltzmann's constant (in joules per kelvin) and $T$ 
is the absolute temperature (in kelvin). In the limit $P_{y}\rightarrow 0$, 
$\epsilon \rightarrow \epsilon_{min}$, which yields $\epsilon \geqslant \epsilon_{min}$. 
Now the efficiency of molecular machine has been defined as the minimum possible 
energy dissipation divided by actual dissipation, i.e.,
\begin{eqnarray*}
\epsilon_{t}\equiv \frac{\epsilon_{min}}{\epsilon} ,
\end{eqnarray*}

\noindent
which is nothing but the isothermal efficiency of a molecular machine 
\citep{Schneider1991a}. Using the channel capacity theorem proposed by Shannon,
Schneider has shown that for a real measurable system, efficiency cannot exceed the 
theoretical limit $\epsilon_t$, mentioned above \citep{Schneider2010}. 
Although, efficiency for both Carnot engine and molecular
machine has been derived using the second law of thermodynamics, the latter is
applicable only for isothermal processes. For more information regarding the concept
of molecular machine and its application we refer to the pioneering work of Schneider
\citep{Schneider1991a, Schneider1991b,Schneider1994,Schneider2010}.
However, to make the present work self contained we briefly review some of the 
notions of information theory developed by \cite{Schneider1991a,Schneider1991b}, 
before going into the application of information theory to our work which we have 
used in analyzing our model.

%%%%% Sequence Logo

\subsection{Sequence logo of primary and secondary binding sites}

Sequence logo is a graphical method which displays the pattern of nucleotides in a set of 
aligned sequences and also provides an idea of affinity to binding sites or preferable binding 
sites for a given sequence \citep{Schneider1990}. In this method occurrence of a base in a 
particular position is 
denoted by the height of that particular base. To signify the conservation at a particular position 
one needs to look at the frequency of occurrence of the base at that position. At this point it is
important to note that \cite{Chauhan2011} have drawn the sequence logo for 
25 DevR directed primary and secondary binding sites.

If one observes both sequence logos for primary and for secondary binding sites, it reveals 
that the logo for primary binding site is more dense than the secondary binding site and hence 
contains more information \citep{Schneider1990}. From this information one can conclude 
that if a promoter contains both the primary and the secondary binding site then the primary 
binding site
majorly controls the transcription which is strongly supported by the expression of different 
mutants. For example, the promoter for Rv1738 gene contains four binding sites, two primary 
(one proximal and one distal) and two secondary \citep{Chauhan2011}. If proximal primary 
binding site is mutated (pBmutD3 following \cite{Chauhan2008b}) the 
expression decreases remarkably (1\% of that of wild type expression). At the same time if 
the proximal secondary binding site is mutated (pBmutD2 following  
\cite{Chauhan2008b}), expression level decreases but 30\% of that of wild type expression 
still persists. Interestingly, when the distal primary binding site is mutated (pBmutD1 following 
\cite{Chauhan2008b}) the expression level remains almost like the wild 
type. Here it should be noted that for DevR regulon, the distance between the transcription start 
point and binding site is very important. For a particular binding site if that distance is large then
it has little contribution to the transcription irrespective of its being primary or secondary, which 
is supported by the expression of the mutant pBmutD1.

What makes the primary sites so different from the secondary sites so that it has control 
over transcription? The sine wave representing the accessibility of a face of DNA (B-form, 
10.6 bases of helical pitch) with the major groove centered at positions 4 and 14.6 
\citep{Schneider1991a,Schneider1991b}. Sequence conservation peak (above 1 bit) at 
positions 4, 5 and 7 and a 10.6 base spacing suggest that DevR makes contact in two 
consecutive major groove through those positions. Hence, these highly conserved positions 
play a major role in binding which is clear from the EMSA result \citep{Chauhan2011}. At the 
same time if one analyzes the logo of the secondary binding sites following the same procedure, 
one finds that there are no such conservation at those positions and hence binding is not so 
strong that ultimately affects the transcription.

From sequence logo one can also judge the DNA bending ability which is an important but 
common structural aspect during transcription. The logo of 120 Fis binding sites shows high 
G and C conservation at $\pm$7 \citep{Shultzaberger2007} so direct contact to major groove 
occurs via these positions. But 
as these positions are close to each other it is difficult to match the D helics into the major groove
properly unless DNA bending occurs. At positions $\pm$4, $\pm$3 and $\pm$2 (central region) 
the logo shows mostly A or T conserved which means either direct minor groove contacts or with 
bending into the minor groove \citep{Schneider2001}. So it may happens that Fis first contacts 
the sequence and bending occurs after that. Similarly the logo of DevR primary binding sites 
contain high conservation at positions $\pm$3, $\pm$5, $\pm$7 (G and C rich) and the central 
region $\pm$1 is A and T rich. In logo the conservation at position $\pm$1 is not so high (just 
greater than 0.5) compared to that of $\pm$3, $\pm$5 and $\pm$7 positions (greater than 1).
If one observes the EMSA mutated at central positions (M-9+9) by C and G, the binding affiny 
vanishes completely (see Fig.~5 of \cite{Chauhan2011}), but this should not be the case as 
conservation at these positions 
are not so high. So one may conclude here that, similar to the previous case discussed, DevR 
binds first to the sequences and bending happens after that. But this is a theoretical prediction 
only, actual scenario is definitely very complex and in a real cellular environment several factors
might play their role which are yet to be verified experimentally. The outcome of the above 
discussion is very important and one should be aware of these facts while making predictions 
for the expression level of different mutants.

%%%%% Ri Rsequence and R frequency

\subsection{ R$_{sequence}$ and R$_{frequency}$}

Transcription factor may bind to many sequences with different affinities. When it binds to a 
specific site, it gains some information. This leads to a natural question, what controls the 
affinity of a transcription factor to a specific binding site? Affinity towards a binding site is 
directly related to the information content of that particular sequence. The high affinity 
binding sites have a greater probability of stabilising the transcription initiation complex 
compared to the low affinity binding sites and thus directly regulates the degree of a 
particular gene expression.

The information of a binding site can be computed by summing the information of each base 
positions of a sequence \citep{Schneider1986}. This is usually done by creating a weight matrix. 
The ri program of Delila was used to create weight matrix by using 25 primary sequences (see 
supplementary information of \cite{Chauhan2011}). The information thus calculated allows 
one to compare between the affinity for two particular binding sites and helps to measure the 
binding energies as well. If one observes the information content of the primary and the secondary 
binding sites carefully it reveals that the primary binding sites generally have more information 
content than the secondary one which again justifies the importance of the primary binding sites 
over the secondary binding sites in connection to the control of particular gene expression.

R$_{frequency}$ depends upon the number of sites and size of  the whole genome 
\citep{Schneider1991a,Schneider1991b}. It is a fixed number which counts the minimum
number of bits required by a protein to bind to a specific site. when this minimal criterion is 
fullfilled, binding takes place on a particular site. The genome of \emph{M. Tuberculossis} is 
$4.6 \times 10^6$ bp long. When a protein comes to bind, it can bind in two possible orientations 
at each base pair. So if a protein wants to bind to 18 sites, it has to choose them from the twice 
$4.6 \times 10^6$ possible binding sites. Hence, the minimum number of binary choices needed 
is R$_{frequency}$ = $\log_2$ ($2 \times 4 \times 10^6/18$) $\approx$ 18.72  bits per site. If one
observes the ratio of R$_{sequence}$/R$_{frequency}$ of the sequences of T7 promoters in 
bacteriophage, it is close to 2. It has been proven experimentally that T7 RNA polymerase only 
uses half of the conserved pattern. In \emph{incD} the ratio is near 3, so at least three proteins 
can bind independently. Most of the systems (including ours too) have the ratio near to 1, that 
means there is just enough pattern at ribosome binding sites (R$_{sequence}$) for them to be 
found in the genetic material of the cell (R$_{frequency}$).

%%%%% Information and Energy

\subsection{Information and Energy}

From the aforesaid discussion we have learned that by exchanging information, protein can 
bind to DNA. So the natural question arises: Is information related to binding energy? Before 
going into the detailed discussion we explore the relation between energy and information.

From the Second Law of Thermodynamics we know the Clausius inequality as
\begin{equation}
d S \geqslant \frac{d Q}{T} .
\end{equation}

\noindent Here $S$ is the total entropy of system, $T$ is the absolute temperature and 
$Q$ is the heat. The protein binding process is an isothermal process and the temperature
remains same immediately after binding. Integration of the above equation, keeping $T$ 
constant yields
\begin{equation}
\Delta S \geqslant \frac{q}{T} .
\end{equation}

\noindent Using the concepts of statistical mechanics one can write the Boltzmann-Gibbs 
entropy of a system as
\begin{equation}
S \equiv -k_B \sum_{i=1}^{\Omega} p_i \ln p_i ,
\end{equation}

\noindent where $k_B$ is the Boltzmann constant, $\Omega$ is the number of possible 
microstates of the system, $p_i$ is the probability of the $i$-th microstate out of $\Omega$
and $\sum_{i=1}^{\Omega} p_i = 1$ for $p_i \geqslant 0$. Following \cite{Shannon1948} 
one can write the uncertainty in each of the microstates as,
\begin{equation}
H \equiv - \sum_{i=1}^{\Omega} p_i \log_2 p_i .
\end{equation}

\noindent Combining Eqs.~(48-49) one can write using $\log_2 (x) = \ln (x) / \ln (2)$,
\begin{equation}
S = k_B \ln(2) H .
\end{equation}

\noindent
The decrease in entropy for an operating machine can be written as
\begin{equation}
\Delta S =S_{after} - S_{before} ,
\end{equation}

\noindent which leads to the following uncertainty in the machine as
\begin{equation}
\Delta H = H_{after} - H_{before} .
\end{equation}

\noindent Now combining Eqs.~(50-52)  one can write 
\begin{equation}
\Delta S = k_B \ln(2) \Delta H .
\end{equation}

\noindent The information gain $R$ by a machine takes place due to decrease in uncertainty
\citep{Shannon1948}, hence one can write
\begin{equation}
R \equiv - \Delta H ,
\end{equation}

\noindent which yields the relation
\begin{equation}
\Delta S = - k_B \ln(2) R .
\end{equation}

\noindent Eq.~(55) shows how the decrease in entropy of a molecular machine is directly 
related to the information that it gains during an operation.
Now substituting Eq.~(55) in Eq.~(47) we get the following inequality
\begin{equation}
k_B T \ln(2) \leqslant \frac{- q}{R} .
\end{equation}

\noindent Eq.~(56) shows how the information is related to heat dissipated ($-q$) during 
an operation. So if a molecular mechine gains 1 bit of information ($R=1$) then minimum 
amount of heat dissipated by the machine is
\begin{equation}
\epsilon_{min}=k_B T \ln(2) \quad \text {(joules per bit)} .
\end{equation}

A protein binds to different sites of a DNA according to its affinity towards that site, so 
protein-DNA dissociation constant, $K_D$ (ratio of rate of association, $k_b$ and 
rate of dissociation, $k_u$), varies with sequence. 
If one thinks from the molecular aspect it is clear that the rate of protein-DNA 
binding depends upon the diffusion rate of the protein. As it can bind to specific as well as to 
non-specific sites one can conclude that apparently the `on' rate is independent of binding 
sequence. As we have discussed previously that a machine should gain some information 
during a successful operation. Similarly, if a protein binds to a non-specific site having lack of 
information then unbinding process is equally probable from that site and hence the 
non-specific site cannot hold the protein to itself. 
But exactly opposite phenomena happens when protein binds to a specific site containing 
the proper information and hence the protein-DNA initiation complex is stabilised and gets 
ready for transcription. From the aforesaid discussion one can conclude that information of 
a binding site ($R_{i}$) is linearly related to the logarithm of `off' rate. But is it really true that 
information has no relation with the $k_b$? To answer this \cite{Shultzaberger2007} have 
shown that $k_b$ (or $K_{on}$ according to \cite{Shultzaberger2007}) is not completely 
independent of information.

Information is related to Gibbs Free energy by a version of Second Law of Thermodynamics
\citep{Berg1987,Berg1988,Barrick1994}
\begin{equation}
R_i \propto -\Delta G .
\end{equation}

\noindent On the other hand Gibbs free energy is related to the dissociation constant via
the relation
\begin{equation}
\Delta G \propto \log{K_D} ; \quad  K_D=\frac{k_u}{k_b} .
\end{equation}

%%%%%%%%%% Figure 15

%\begin{figure}[!t]
%\begin{center}
%\includegraphics[width=0.75\linewidth,angle=0]{ri.eps}
%\includegraphics[width=0.75\linewidth,angle=0]{ri.pdf}
%\end{center}
%\caption{Plot of individual information with the logarithmic values of model parameters, $k_b$,  
%$k_u$ and their ratio  $K_{D}$. Solid squares are the logarithm of those parameters and straight 
%lines are the linear fit. This plot shows that logarithmic values of model parameters and the individual 
%information are in linear relationship which can be explained from the information theory .
%}
%\label{inform}
%\end{figure}

\noindent Relating Eqs.~(58-59) yields
\begin{equation}
R_i \propto -\log \frac{k_u}{k_b} .
\end{equation}

\begin{figure}[!t]
\begin{center}
\includegraphics[width=0.75\linewidth,angle=0]{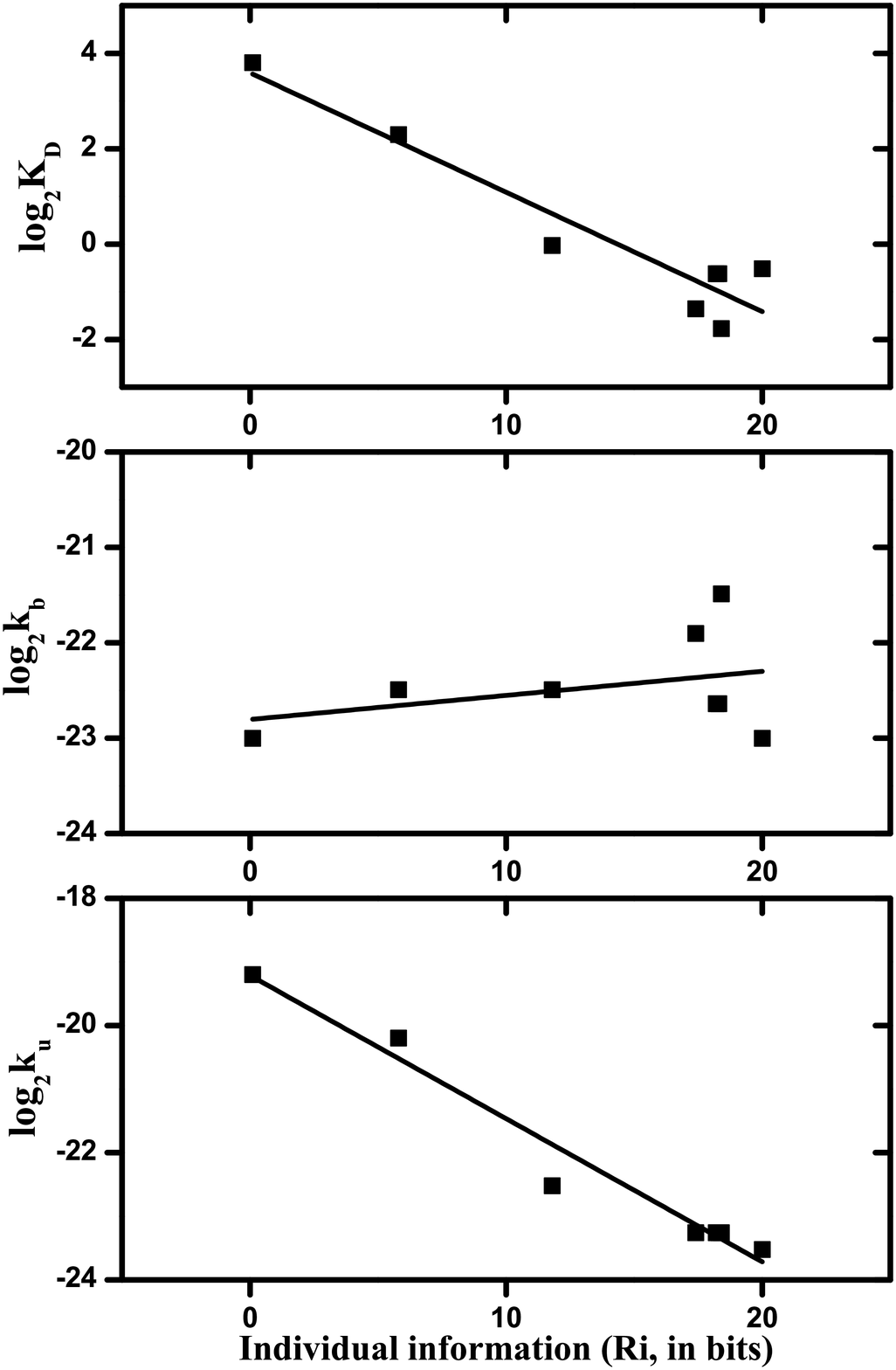}
\end{center}
\caption{Plot of individual information with the logarithmic values of model parameters, 
$k_b$,  $k_u$ and their ratio  $K_{D}$. Solid squares are the logarithm of the parameters 
and straight lines are the linear fit. This plot shows that logarithmic values of model parameters 
and the individual information are in a linear relationship.
}
\label{inform}
\end{figure}

\noindent So the relation between information and $k_u$ (or $K_{off}$ according to
\cite{Shultzaberger2007}) is negatively proportional which means more the information 
the sequences have, it is more difficult to destabilise the transcription 
initiation complex. Thus, according to the information theory, information has linear relationship 
with both the quantity $K_{D}$ and $k_u$ with negative slope. From our model parameter
value we observe that such linear relationship holds good pretty well as discussed above
(see Fig.~\ref{inform}). Interestingly the $k_b$ rate remains almost constant as protein binds
frequently to a binding site irrespective of its affinity to that particular site
\citep{Das2005,Kim1987,Linnell2004,Schaufler2003,Shultzaberger2007}, 
which is also evident in our case (see the middle panel of Fig.~\ref{inform}).

%%%%% Molecular efficiency

\subsection{Molecular efficiency}

The term `efficiency'  was first introduced in classical thermodynamics in the context of a 
heat engine \citep{Callen1985,Guemez2002,Jaynes2003} which operates between two 
reservoirs at temperature \emph{$T_{hot}$} and \emph{$T_{cold}$},
\begin{equation}
\eta = \frac {T_{hot}-T_{cold}}{T_{hot}} .
\end{equation}

\noindent But this equation is not valid in the biological context because to get 70\% 
efficiency \emph{$T_{hot}$} and  \emph{$T_{cold}$} need to be 1000k and 300k, 
respectively, which is lethal for biological systems \citep{Jaynes1988}. Another reason 
due to which it is not applicable in biological systems is the isothermal nature of most of 
the biological processes. As a result of which one needs an expression for efficiency for 
isothermal processes \citep{Schneider1991b}. From information theoretic point of view 
one can measure the isothermal efficiency when a protein gets bound to a specific site 
by the relation
\begin{equation}
\epsilon_{r}=\frac {R_{sequence}}{R_{energy}} ,
\end{equation}

\noindent where we have defined $R_{sequence}$ previously. Here $R_{energy}$ is 
logarithm of $K_{spec}$ where $K_{spec}$ is the ratio of specific and nonspecific binding 
at a particular site \citep{Schneider2010}
\begin{equation}
R_{energy}=\log_2 K_{spec} \quad \text{where} \quad K_{spec}=\frac {k_{s}}{k_{n}} .
\end{equation}

\noindent Here $k_s = 1/K_D$ ($K_D$ values for different binding sites are listed
in Table~1). For the binding site P1 which is in the intergenic region of narK2-Rv1738 
the $K_{D}$ value is $0.697 \times 10^{-9}$ M. It is important to mention that the nonspecific 
binding energy is not known for this system. Considering 
$\log_2 k_n = 0$ (as the nonspecific binding energy is not known), we find $R_{energy} 
\approx \log_2 k_{s} = 30.41$ (bits per site). Henceforth $\epsilon_{r}=20/30.41 \simeq 0.66$, 
where $R_{sequence} \approx 20$. So according to our mathematical model efficiency of 
this system is 66\% and if one calculates the efficiency of other primary binding sites by 
following the same procedure one will find that all $\epsilon_r$ values are around 60-65\% 
efficient. This is pretty close to the maximum limit of isothermal efficiency of  70\% as reported 
by \cite{Schneider2010}. Note that there are many systems like EcoRI, RepA, etc., which has 
the efficiency close to this maximum limit.

At this point it is important to mention that for primary binding sites the efficiency is 
quite good but if one calculates the same for secondary binding sites the efficiency will
be quite low as many of them have low $R_{sequence}$ value. This finding is another 
justification of why the secondary binding sites have lower contribution to transcription 
compared to the primary binding sites. Beside this, for Rv3134c, both primary and 
secondary binding sites have similar molecular efficiency which again raise the question 
that whether the construct is P-P or P-S?

%%%%% Conclusion

\section{Conclusion}

The DevRS two component system of \emph{M. tuberculosis} is responsible for its 
dormancy in host and becomes operative under hypoxic condition. It is experimentally
known that phosphorylated DevR controls expression of several downstream genes 
in a complex manner. To understand the mechanism of DevR mediated downstream gene
regulation we have developed a theoretical model based on the elementary kinetics
of DevR-promoter interaction. The kinetic model we have developed is 
efficient in describing behavior of some DevR regulated genes. To this end, we have 
chosen four DevR controlled genes and have shown that our proposed model can qualitatively
generate the gene expression profile of the wild type strain and some novel mutants that are
impaired in DevR binding site. The DevR regulated promoter sites have a definite 
pattern of construction which contains one stronger binding site (primary sites) and nearly 
located relatively weaker binding site (secondary site). From construction of the binding sites 
it seems that primary binding sites majorly control the gene expressions mechanism with
a little contribution from the secondary binding sites. Through modeling, we have shown that
when both sites (primary as well as secondary) impart a co-operative contribution towards 
the DevR binding mechanism, effect of the secondary binding site is not negligible.
This phenomenon can also be understood from expression profile of some mutants we 
have predicted in the present study. Keeping this binding pattern in mind we have thus
proposed a generalized mechanism which can be applied to understand the temporal
profile for any DevR regulated genes. 
From the information theoretical analysis we have seen that the primary binding sites 
contain more information than the secondary binding sites which justify the above 
mentioned mechanism of the preference of DevR towards primary binding sites over 
secondary binding sites. From information theory it is known that the binding rate constants 
are in a linear relationship with the individual information of the binding sites 
\citep{Schneider2010}. The parameter sets we have used for modeling could generate 
this linear relation predicted by information theory (see Fig.~\ref{inform}). Another important 
aspect information theory predicts is the molecular efficiency.
Using information theory it can be shown that maximum limit of isothermal efficiency is 
70\% \citep{Schneider2010}. From our model we have calculated the molecular efficiency 
of the system and have shown that it is close to the maximum limit of isothermal efficiency.
Thus in totality, the proposed model could recapture the experimental aspects of DevR
mediated gene expression and helps one to understand the phenomenon from
information theoretic point of view.
We hope that our theoretical model and the subsequent analysis will inspire more 
experiments in coming days to address other critical issues of DevR regulatory networks 
that are yet to be explored. Information from this new experimental data will help one to 
build more detailed model in future.

\begin{acknowledgements}
We express our sincerest gratitude to Jaya S Tyagi and Thomas D Schneider for 
stimulating discussions and suggestions.
AB acknowledges CSIR, Government of India, for a research fellowship (09/015(0375)/2009-EMR-I). 
SB acknowledges support from Centre of Excellence (CoE) at Bose Institute, Kolkata, 
supported by DBT, Government of India.
AKM acknowledges UGC, Government of India, for a research fellowship
(UGC/776/JRF(Sc)). 
SKB acknowledges support from Bose Institute through Institutional Programme 
VI - Development of Systems Biology.
\end{acknowledgements}

\appendix

\section*{Appendix}

\noindent DevR:

\begin{equation}
\frac{d[R_{p}]}{dt}=k_{srp}-k_{drp} [R_{p}] .
\end{equation}

\noindent {\it Rv3134c}:

\begin{eqnarray}
\frac{d[P^*]}{dt}&=&k_{b1}[P][R_{p}] - k_{u1}[P^*], \\
\frac{d[S^*]}{dt}&=&k_{b2}[S][R_{p}] - k_{u2}[S^*], \\
\frac{d[mGFP_{4c}]}{dt}&=&k_{sm1}[P^*]+ k_{sm2}[S^*]+ k_{sm3}[P^*][S^*] \nonumber \\
&& - k_{dm}[mGFP_{4c}], \\
\frac{d[GFP]}{dt}&=&k_{sg}[mGFP_{4c}] - k_{dg}[GFP] .
\end{eqnarray}

\noindent {\it hspX}:

\begin{eqnarray}
\frac{d[P1^*]}{dt}&=&k_{b3}[P1][R_{p}] - k_{u3}[P1^*], \\
\frac{d[P2^*]}{dt}&=&k_{b4}[P2][R_{p}] - k_{u4}[P2^*], \\
\frac{d[S^*]}{dt}&=&k_{b5}[S][R_{p}] - k_{u5}[S^*], \\
\frac{d[mGFP_{px}]}{dt} &= &k_{sm4}[P1^*]+ k_{sm5}[P2^*]+ k_{sm6}[S^*]  \nonumber \\
&& + k_{sm7}[P1^*][P2^*][S^*] - k_{dm}[mGFP_{px}], \\
\frac{d[GFP]}{dt}&=&k_{sg}[mGFP_{px}] - k_{dg}[GFP].
\end{eqnarray}

\noindent {\it narK2-Rv1738}:

\begin{eqnarray}
\frac{d[P1^*]}{dt}&=&k_{b6}[P1][R_{p}] - k_{u6}[P1^*], \\
\frac{d[P2^*]}{dt}&=&k_{b7}[P2][R_{p}] - k_{u7}[P2^*], \\
\frac{d[S1^*]}{dt}&=&k_{b8}[S1][R_{p}] - k_{u8}[S1^*], \\
\frac{d[S2^*]}{dt}&=&k_{b9}[S2][R_{p}] - k_{u9}[S2^*], \\
\frac{d[mGFP_{K2}]}{dt} &=&k_{sm8}[P1^*]+ k_{sm10}[P2^*]+ k_{sm12}[S1^*] \nonumber \\
&&+ k_{sm14}[S2^*]+ k_{sm16}[P1^*][S1^*] \nonumber \\
&& + k_{sm18}[P2^*][S2^*] - k_{dm}[mGFP_{K2}], \\
\frac{d[GFP]}{dt} &=& k_{sg}[mGFP_{K2}] - k_{dg}[GFP], \\
\frac{d[mGFP_{38}]}{dt}&=&k_{sm9}[P1^*]+ k_{sm11}[P2^*]+ k_{sm13}[S1^*] \nonumber \\
&&+ k_{sm15}[S2^*]+ k_{sm17}[P1^*][S1^*] \nonumber \\
&& + k_{sm19}[P2^*][S2^*]  - k_{dm}[mGFP_{38}], \\
\frac{d[GFP]}{dt}&=&k_{sg}[mGFP_{38}] - k_{dg}[GFP].
\end{eqnarray}

%%%%%%%%%% References

\newpage

%%%%% Table 1

\begin{table}
\label{tab1}
\begin{center}
\begin{tabular}{lllll}
\hline
Promoter & Site & $k_{b} \times 10^{-7}$ & $k_{u} \times 10^{-7}$ & $K_D$ \\
                  &         & (nM$^{-1}$ s$^{-1}$)    & (s$^{-1}$)                        & (nM) \\
\hline
Rv3134c & P & 1.70 & 1.0 & 0.588 \\
                 & S & 1.70 & 1.0 & 0.588 \\
\hline
\textit{hspX} & P1 & 3.413  & 1.0 & 0.293 \\       
                      & P2 & 1.365  & 8.33 & 6.102 \\    
                      & S   &  1.706  & 16.67 & 9.771 \\              
\hline
\textit{narK2}-Rv1738 & P1 & 1.194 & 0.833 & 0.697 \\  
                                        & P2 & 2.559 & 1.0 & 0.391 \\      
                                        & S1 & 1.194 & 16.6 & 13.903 \\      
                                        & S2 & 1.70 & 1.66 & 0.976 \\      
\hline
\end{tabular}
\end{center}
\caption{List of binding $k_{bi}$ ($i=1-9$) and unbinding $k_{ui}$ ($i=1-9$) constants
for the promoters Rv3134c, \textit{hspX} and \textit{narK2}-Rv1738. The corresponding
$K_D$ ($=k_u/k_b$) value for each binding site are also given.
}
\end{table}

%%%%% Table 2

\begin{table}
\label{tab2}
\begin{center}
\begin{tabular}{lll}
\hline
Parameter & Value & Description \\
\hline
$k_{srp}$ & $4.07 \times 10^{-3}$ nM s$^{-1}$ & Synthesis of $R_p$ \\
$k_{drp}$ & $1.66 \times 10^{-5}$ s$^{-1}$ & Degradation of $R_p$ \\
$k_{sm1}$ & $2.44 \times 10^{-4}$ nM s$^{-1}$ & Synthesis of $mGFP_{4c}$ from $P^*$ \\
$k_{sm2}$ & $2.44 \times 10^{-4}$ nM s$^{-1}$ & Synthesis of $mGFP_{4c}$ from $S^*$ \\
$k_{sm3}$ & $4.90 \times 10^{-3}$ nM s$^{-1}$ & Synthesis of $mGFP_{4c}$ from $P^*S^*$ \\
$k_{dm}$ & $8.33 \times 10^{-4}$ s$^{-1}$ & Degradation of $mGFP_{4c}$ \\
$k_{sm4}$ & $4.22 \times 10^{-3}$ nM s$^{-1}$ & Synthesis of $mGFP_{px}$ from $P1^*$ \\
$k_{sm5}$ & $8.13 \times 10^{-4}$ nM s$^{-1}$ & Synthesis of $mGFP_{px}$ from $P2^*$ \\
$k_{sm6}$ & $1.29 \times 10^{-4}$ nM s$^{-1}$ & Synthesis of $mGFP_{px}$ from $S^*$ \\
$k_{sm7}$ & $3.25 \times 10^{-3}$ nM s$^{-1}$ & Synthesis of $mGFP_{px}$ from $P1^*P2^*S^*$ \\
$k_{dm}$ & $8.33 \times 10^{-4}$ s$^{-1}$ & Degradation of $mGFP_{px}$ \\
$k_{sm8}$ & $5.20 \times 10^{-4}$ nM s$^{-1}$ & Synthesis of $mGFP_{K2}$ from $P1^*$ \\
$k_{sm9}$ & $1.62 \times 10^{-4}$ nM s$^{-1}$ & Synthesis of $mGFP_{38}$ from $P1^*$ \\
$k_{sm10}$ & $1.62 \times 10^{-5}$ nM s$^{-1}$ & Synthesis of $mGFP_{K2}$ from $P2^*$ \\
$k_{sm11}$ & $1.13 \times 10^{-3}$ nM s$^{-1}$ & Synthesis of $mGFP_{38}$ from $P2^*$ \\
$k_{sm12}$ & $1.62 \times 10^{-5}$ nM s$^{-1}$ & Synthesis of $mGFP_{K2}$ from $S1^*$ \\
$k_{sm13}$ & $1.62 \times 10^{-5}$ nM s$^{-1}$ & Synthesis of $mGFP_{38}$ from $S1^*$ \\
$k_{sm14}$ & $1.62 \times 10^{-5}$ nM s$^{-1}$ & Synthesis of $mGFP_{K2}$ from $S2^*$ \\
$k_{sm15}$ & $7.00 \times 10^{-4}$ nM s$^{-1}$ & Synthesis of $mGFP_{38}$ from $S2^*$ \\
$k_{sm16}$ & $4.88 \times 10^{-4}$ nM s$^{-1}$ & Synthesis of $mGFP_{K2}$ from $P1^*S1^*$ \\
$k_{sm17}$ & $1.62 \times 10^{-3}$ nM s$^{-1}$ & Synthesis of $mGFP_{38}$ from $P1^*S1^*$ \\
$k_{sm18}$ & $3.25 \times 10^{-4}$ nM s$^{-1}$ & Synthesis of $mGFP_{K2}$ from $P2^*S2^*$ \\
$k_{sm19}$ & $6.10 \times 10^{-3}$ nM s$^{-1}$ & Synthesis of $mGFP_{38}$ from $P2^*S2^*$ \\
$k_{dm}$ & $8.33 \times 10^{-4}$ s$^{-1}$ & Degradation of $mGFP_{K2}$ \\
$k_{dm}$ & $8.33 \times 10^{-4}$ s$^{-1}$ & Degradation of $mGFP_{38}$ \\
$k_{sg}$ & $6.66 \times 10^{-4}$ nM s$^{-1}$ & Synthesis of GFP \\
$k_{dg}$ & $1.67 \times 10^{-5}$ s$^{-1}$ & Degradation of GFP \\
\hline
\end{tabular}
\end{center}
\caption{List of kinetic parameters (with values) used in the model. Note that, degradation
constant ($k_{dm}$) for all four $mGFP$-s ($mGFP_{4c}$, $mGFP_{px}$, $mGFP_{K2}$ 
and $mGFP_{38}$) have been considered to be same.
}
\end{table}

\newpage

%%%%% Figure 1

%%%%% Figure 2

%%%%% Figure 3

%%%%% Figure 4

%%%%% Figure 5

%%%%% Figure 6

%%%%% Figure  7

%%%%% Figure 8

%%%%% Figure 9

%%%%% Figure 10

%%%%% Figure 11

%%%%% Figure 12

%%%%% Figure 13

%%%%% Figure 14

%%%%%%%%%% Figure 15

%%%%%%%%%% Figure 16

\end{document}